\documentclass[fleqn,usenatbib]{mnras}
\usepackage{newtxtext,newtxmath}
\usepackage[T1]{fontenc}
\usepackage{ae,aecompl}

\DeclareRobustCommand{\VAN}[3]{#2}
\let\VANthebibliography\thebibliography
\def\thebibliography{\DeclareRobustCommand{\VAN}[3]{##3}\VANthebibliography}

\usepackage{graphicx}	% Including figure files
\usepackage{amsmath}	% Advanced maths commands
\newcommand{\kms}{\,km\,s$^{-1}\,$}   % kilometres per second
\newcommand{\Msun}{M$_{\sun}$}     % solar mass
\newcommand{\Lsun}{L$_{\sun}$}     % solar luminosity
\newcommand{\Rsun}{R$_{\sun}$}     % solar luminosity
\newcommand{\Msuny}{M$_{\sun}$\,yr$^{-1}$}      % solar mass per year
\newcommand{\Hal}{H${\small\alpha}\,$}     % H alpha
\newcommand{\DNa}{D Na\,{\sc i}\,}     % DNaI
\newcommand{\Teff}{$T_{\textrm{eff}}$~}     %Effective temperature
\newcommand{\vsini}{$v \sin i$}     %vsini

\title[Density streams in the disc winds of Classical T Tauri stars]{Density streams in the disc winds of Classical T Tauri stars}

\author[P. P. Petrov et al.]{
P. P. Petrov,$^{1}$\thanks{E-mail: petrov@craocrimea.ru}
K. N. Grankin,$^{1}$
E. V. Babina,$^{1}$
S. A. Artemenko,$^{1}$
M.M. Romanova,$^{2}$
S. Yu. Gorda,$^{3}$
\newauthor {A. A. Djupvik,$^{4,5}$ J. F. Gameiro,$^{6,7}$}
\\
$^{1}$Crimean Astrophysical Observatory, p/o Nauchny, 298409, Republic of Crimea\\
$^{2}$Cornell University, Ithaca, NY 14853, USA\\
$^{3}$Ural Federal University, 51, Lenin av., Ekaterinburg, Russia, 620000\\
$^{4}$Nordic Optical Telescope, Rambla Jos\'{e} Ana Fern\'{a}ndez P\'{e}rez, 7, 38711 Bre\~{n}a Baja, Spain\\
$^{5}$Department of Physics and Astronomy, Aarhus University, Ny Munkegade 120, 8000 Aarhus C, Denmark\\
$^{6}$Instituto de Astrofísica e Ci\^{e}ncias do Espa\c{c}o, Universidade do Porto, CAUP, Rua das Estrelas, PT4150-762 Porto, Portugal\\
$^{7}$Departamento de Física e Astronomia, Faculdade de Ci\^{e}ncias, Universidade do Porto, Rua do Campo Alegre 687, PT4169-007 Porto, Portugal\\
}

\date{Accepted XXX. Received YYY; in original form ZZZ}

\pubyear{2023}

\begin{document}
\label{firstpage}
\pagerange{\pageref{firstpage}--\pageref{lastpage}}
\maketitle
\setcitestyle{note Paper I}
\defcitealias{Petrov2019}{Paper I}

\begin{abstract}
Spectral and photometric variability of the  Classical  T Tauri stars RY Tau and SU Aur from 2013 to 2022  is analyzed. We find that in SU Aur the  \Hal line's flux at radial velocity RV = $-50 \pm 7$ \kms varies with a period P = $255 \pm 5$ days.  A similar effect previously discovered in RY Tau is confirmed with these new data: P = 21.6 days at RV =$ -95 \pm 5$ \kms. In both stars, the radial velocity of these variations, the period, and the mass of the star turn out to be related by Kepler's law, suggesting  structural features on the disc plane orbiting  at radii of 0.2 AU in RY Tau and 0.9 AU in SU Aur, respectively. Both stars have a large inclination of the accretion disc to the line of sight  - so that the line of sight passes through the region of the disc wind.  We propose there is an azimuthal asymmetry in the disc wind, presumably in the form of  'density streams', caused by substructures of the accretion disc surface. These streams cannot dissipate until they go beyond the Alfven surface in the disc's magnetic field. These findings open up the possibility to learn about the structure of the inner accretion disc of CTTS on scales less than 1 AU and to reveal the orbital distances related to the planet's formation.

\end{abstract}


%%%%%%%%%%%%%%%%% BODY OF PAPER %%%%%%%%%%%%%%%%%%

\section{Introduction}

The Classical T Tauri Stars (CTTSs) are commonly recognized as the counterparts of the young Sun, allowing us to study the solar system's past evolution.  The formation of a planetary system around a young star occurs simultaneously with the formation of the star itself - when the central star and the surrounding protoplanetary disc are still hidden from the observer by the gas-dust shell  \citep{Nixon2018}. By the time we see a star as a CTTS, there are already possible indirect  signatures of planets in the surrounding disc. However, detecting these planets is quite difficult. 
The classical method of precision measurements of radial velocities is hardly applicable for CTTSs: as a result of magnetospheric accretion, hot spots form on the stellar surface, distorting the photospheric spectrum  \citep[e.g.][]{Petrov2011}. Unlike long-lived cool spots, these hot spots can change on a time scale of a few days, resulting in jittering the measured radial velocity. Although thousands of exoplanets are currently discovered and their properties and migrations are being investigated, we still have very limited information about planets around CTTSs (see \citealt{Grandjean2021} and references therein).  

In this article, we consider the possibility of detecting emerging planets in the inner accretion disc by their influence on the disc wind. 

The wind properties in CTTS  are usually inferred from the kinematics of spatially unresolved, blueshifted forbidden emission lines including [O\,{\sc i}], [S\,{\sc ii}], [N\,{\sc ii}], [Fe\,{\sc ii}], [Ne\,{\sc ii}], as well as P Cygni profiles in strong permitted emission lines, such as He\,{\sc i} 10830 \AA \,(see review by \citealt{Pascucci2022} and references therein). Analysis of the FUV spectra  of CTTS showed that the mass loss is concentrated in the innermost region of the disc, < 1-2 AU \citep{Xu2021}.

In the visible spectral region the most obvious signature of the wind is the variable blue-shifted depression in the \Hal emission profile. The radial velocity of the blue-shifted depression depends on the wind velocity and the inclination angle of the disc axis to the line of sight.

\begin{table*}
\label{tab:tab1}
\centering
\caption{The basic parameters of SU Aur and RY Tau}
\begin{tabular}{cccccccc} 
\hline
Star      & Sp. type    &  \Teff   & R/\Rsun  & L/\Lsun  & M/\Msun   & $\dot{M}$ & inner disc inclination \\
            &                   &      K     &      		&          &                  & $10^{-8}$ \Msuny       &   degrees    \\
\hline
SU Aur  &  G1            & 5945  & $2.6 \pm 0.4$   & $7.8 \pm 1.2$ &  $1.7 \pm 0.2$ & $0.5 - 0.6 (\pm 0.4)$ & $50.9 \pm 1$  \\
RY Tau  &  G1            & 5945  & $2.9 \pm 0.4$    &  $9.6 \pm 0.5$ & $2.0 \pm 0.3$ & $6.4 - 9.1 (\pm 4.9)$ & $60.0 \pm 1$\\
\hline
\multicolumn{8}{l}{Stellar parameters:  \citet{Calvet2004}.}\\
\multicolumn{8}{l}{Inner disc inclinations: \citet{Labdon2019} (SU Aur); \citet{Perraut2021} (RY Tau).}\\

\end{tabular}
\end{table*}

However, when observing the variability of the \Hal line formed in the CTTS wind, it is difficult to identify the type of wind responsible for the changes in the line profile. Different types of outflow can start at the boundary between the stellar magnetosphere and the inner accretion disc. The disc wind, accelerated by the magneto-centrifugal force, emanates from the disc surface \citep{Blandford1982}. Besides, the X-wind \citep{Shu1994}, conical wind \citep{Romanova2009,Kurosawa2012}, magnetospheric ejections \citep{Zanni2013}, and magnetic propeller were discussed (see \citealt{Romanova2015,Romanova2018} for review).

Emission line profiles provide information about the dynamics of gas flows around a CTTS. The formation of various types of the \Hal emission profiles in spectra of CTTS was considered by \citet{Kurosawa2006} within models of the magnetosphere and the disc wind. They have shown that at a sufficiently high inclination angle of the line of sight, the main feature in the spectral line is blue-shifted absorption due to the disc wind. A similar result has been obtained while modeling the disc wind in the He\,{\sc i} (10830) spectral line \citep{Kurosawa2011}.

Models of \Hal line profiles formed in the magnetosphere  and the disk wind of CTTSs  were  calculated and compared with the observed ones \citep{Kurosawa2011,Tambovtseva2014}.

Variability of spectral line profiles caused by the gas flows, including the outflows from the boundary between the magnetosphere and the accretion disk, is well documented for some CTTS, e.g.  by \citealt{Sousa2021} (V2129 Oph),  \citealt{Alencar2018}  (LkCa15), \citealt{Bouvier2007a} (AA Tau), \citet{Bouvier2023} (GM Aur), while less is known about variations in the disc wind. 

There is a class of young stars - FUors - where the disc wind dominates over other types of outflow. The prototype of the class, FU Ori, is a  low-mass CTTS with the mass-loss rate in the disc wind $2\times10^{-4}$ \Msuny \citep{Zhu2007}, while in a CTTS the average value is $10^{-9}$ \Msuny  \citep{Bouvier2007b}.    \citet{Herbig2003} analyzed a series of spectra of FU Ori, obtained in 1997 - 1999. They found that the intensity of the P Cyg absorption in the \Hal line varied with a period of 14.8 days. A decade later, \citet{Powell2012} analyzed their series of spectral observations of FU Ori of 2007. They confirmed the existence of the periods and refined its values: 13.5 days. The case of FU Ori shows that periodic variations in the disc wind are real and worth to be investigated.

A long time series of spectroscopic observations of a CTTS may potentially reveal rotationally modulated variations in the wind density at certain radial velocities. Such a Doppler probe may provide information about the inner disc structure. In this article we focus on time variability of \Hal line fluxes at different radial velocities across the broad line profile.

We present the results of spectral and photometric monitoring of  RY Tau and SU Aur, both with large inclinations, so that line of sight intersects the region of the disc winds. The stars are bright, and each has been explored in detail before. The stellar parameters, according to \citet{Calvet2004}, are presented in Table~\ref{tab:tab1}. The stars are relatively massive and may be classified as the intermediate objects between the CTTS and the HAeBe stars.

Our program of spectroscopic and photometric monitoring of RY Tau were started in 2013, then later in 2015 another object, SU Aur, was included.  The first results of this program were published in \citet{Petrov2019}, hereafter referred as  \citetalias{Petrov2019}: from analysis of the variable outflow velocity and the circumstellar extinction of RY Tau we found that the obscuring dust is near the star, at the dust sublimation radius of about 0.2 AU. During events of enhanced outflow, the circumstellar extinction becomes lower. As more data on RY Tau accumulated, we found another effect: the spectral indicators of the accretion and outflows (in \Hal and Na\,{\sc i} lines) vary with a period of about 22 d. The infall and outflow vary in antiphase: an increase of infall is accompanied by a decrease of outflow, and vice versa. We considered two possible interpretations: 1) quasi-periodic magnetohydrodynamics processes at the disc-magnetosphere boundary in the propeller mode, or 2) a massive planet on inclined orbit in the accretion disc \citep{Petrov2021}.

By now we have accumulated 9 seasons of observation of RY Tau and 7 seasons of SU Aur, including  high resolution spectroscopy and photometry. In this paper we investigate variability of the \Hal profile in spectra of RY Tau and SU Aur, with the aim to confirm the periodic variations and find their possible relationship with the structure of the accretion disc and the disc wind.

The paper is organized as follows. In Section~\ref{sec:properties}  we overview the basic properties of RY Tau and SU Aur.  In Section~\ref{sec:observations}  we address briefly the instruments  and observation sites. Then, in Section~\ref{sec:results} we analyze the variability of brightness and \Hal line fluxes in RY Tau and SU Aur. The major part is devoted to the frequency analysis of variability of the \Hal line flux at different radial velocities. Finally, in Section~\ref{sec:discussion}  we discuss the results in terms of the disc wind axial asymmetry and the accretion disc structures.

\section{Basic properties of RY Tau and SU Aur}
\label{sec:properties}

RY Tau and SU Aur are young stars with emission line spectra, showing signatures of outflows and accretion. Both stars possess accretion discs. Excess continuous radiation is present in the far UV spectrum of both stars, larger in RY Tau than in SU Aur. The mass accretion rates (Table~\ref{tab:tab1}) were estimated from the accretion luminosities in the UV spectrum \citep{Calvet2004}.

Inclinations of the \textit {inner} disc were  derived from $ K$-band interferometry:  $60^\circ\pm1^\circ$ in RY  Tau \citep{Perraut2021}  and  $50.9^\circ\pm1^\circ$ in SU Aur \citep{Labdon2019}. The inclination of the \textit {outer} disc of RY Tau, as derived from ALMA observations, is $65^\circ\pm 0.1^\circ$  \citep{Long2019}. $K$-band interferometry of RY Tau by \citet{Davies2020} indicated that the central star is occulted by the disc surface layers close to the sublimation rim. They suggested that the aperiodic photometric variability of RY Tau is likely related  to temporal and/or azimuthal variations in the structure of the disc surface layers.

Interferometric and polarimetric observations of SU Aur revealed  more complicated structures with several dust tails connected to the Keplerian disk, indicating that late accretion events 
can still occur \citep{Ginski2021,Akiyama2019}.

Both stars are rapid rotators. In RY Tau, \vsini \, = $52 \pm 2$ \kms \citep{Bouvier1990,Petrov1999}. The period of axial rotation was not detected directly either in photometric data or in variations of spectral line profiles. Assuming the inclination of the stellar rotational axis $i = 60^\circ$ (the same as in the accretion disc), stellar radius $ 2.9 \pm 0.4$ \Rsun \, and \vsini \,= 52 \kms,  the expected period of the axial rotation  may be about 3 days.

In SU Aur, \vsini \, = $60-66$ \kms \citep{Nguyen2012}. The period of axial rotation within $2.7 - 3.0$ days was estimated from variations in emission line profiles \citep{Giampapa1993,Johns1995} and in the red-shifted absorptions of He\,{\sc i} and Balmer lines \citep{Petrov1996}. 

RY Tau has an extended jet with a few knots of young dynamical ages \citep{St-Onge2008,Agra2009,Skinner2011}.  The jet is wiggled, which may indicate the presence  of an unseen planetary or sub-stellar companion to the star \citep{Garufi2019}. Interferometric images of the protoplanetary disc around RY Tau at millimeter wavelengths did not reveal planets more massive than 5 M$_J$ at distances between 10 and 60 AU \citep{Isella2010}.

Both RY Tau and SU Aur are irregular variables with long photometric records \citep{Herbst1984,Herbst1994,Zajtseva1996}. No stable periods  in the light variations were recorded,  
although  quasi-periodic variations  on different time scales were reported by observers. The most extended series of photometric observations of RY Tau from 1965 to 2000 were analyzed 
by \citet{Zajtseva2010}:  periods of 20.0 and 29.4 days were revealed in the data of 1993 and 1996 respectively and interpreted as probably caused by inhomogeneities in the circumstellar disc. A period of about 23 days in variations of emission line intensities in UV and optical spectrum of RY Tau was reported by \citet{Ismailov2015} and interpreted as due to protostellar objects 
orbiting in the inner accretion disc. 

The light curve of SU Aur is characterized by a nearly constant maximum brightness level with a usually small amplitude of variability but interrupted at times by deep fading episodes
\citep{Grankin2007}. Searches for periodic photometric variations gave conflicting results. Various authors reported finding different periods: 1.76 d \citep{DeWarf2003}, 1.55 d or 2.73 d 
\citep{Herbst1987}, 2.7 d \citep{Strassmeier1997} or their absence \citep{Percy2006,Artemenko2012}. Only analysis of intense space observations performed on the MOST satellite made it possible to reliably detect a low-amplitude periodic process (2.661 days) against the background of a longer and higher-amplitude brightness change \citep{Cody2013}.

All the reported periods in photometric and spectroscopic observations of SU Aur and RY Tau are summarized in Table~\ref{tab:tab2}. More detailed review of the previously reported photometric periods in SU Aur was presented in \citet{Cody2013}. In summary:  the short periods (< 3 days) in SU Aur indicate the axial rotation of the star,  while the longer periods observed in RY Tau are probably related to rotations of  the inner region of the accretion disk. A more detailed overview of the basic properties of RY Tau and SU Aur was presented in \citetalias{Petrov2019}.

\begin{table}
\label{tab:tab2}
\centering
\caption{Reported periods in photometric and spectroscopic observations of SU Aur and RY Tau}
%\scriptsize{
\begin{tabular*}{0.95\columnwidth}{ccccc} 
%\caption{Reported periods in photometric and spectroscopic observations of SU Aur and RY Tau}
\hline
Star & Period, days    &  Method   &Year of obs.  & Reference \\
\hline
SU Aur&  1.55; 2.73       & phot  & 1985 - 1986 & 1  \\
           &   $\sim 1.7$       & phot  & 1995 - 1999 & 2  \\
            &  2.661       & phot  & 2009 - 2010 & 3  \\
            &  2.98       & spec  & 1986 - 1990 & 4  \\
            &   3.0       & spec  & 1986 - 1992 & 5  \\
            &  3.0       & spec  & 1990 - 1994 & 6  \\
            &  2.742       & spec  & 1993 - 1994 & 7  \\
            & 18.7       & phot  & 1981 - 1991 & 8  \\
\hline
 RY Tau &20; 29.4 &  phot & 1993, 1996 & 9 \\
          & 23          & spec & 1975 - 1985 & 10 \\
	&  21.6 	& spec & 2013 - 2021 & 11\\
\hline
\multicolumn{5}{l}{References: 1 - \citet{Herbst1987}, 2 - \citet{DeWarf2003},}\\
\multicolumn{5}{l}{3 - \citet{Cody2013}, 4 - \citet{Giampapa1993},}\\
\multicolumn{5}{l}{5 - \citet{Johns1995}, 6 - \citet{Petrov1996},}\\
\multicolumn{5}{l}{7 - \citet{Strassmeier1997}, 8 - \citet{Gahm1993}, 9 - \citet{Zajtseva2010},}\\
\multicolumn{5}{l}{10 - \citet{Ismailov2015}, 11 - \citet{Petrov2019,Petrov2021}.}\\
\end{tabular*}
%}
\end{table}

\section{Observations}
\label{sec:observations}

A major part of our spectral data was obtained with an echelle spectrograph at the 2.6 meters Shajn reflector of the Crimean Astrophysical Observatory (CrAO) and with a fiber-fed echelle spectrograph at the 1.21-meter telescope of Kourovka Observatory of the Ural Federal University. The spectral resolutions $\lambda/\Delta\lambda$ were 27000 and 15000 correspondingly. Two spectral orders, including \Hal and \DNa lines, were extracted from the echelle frame for analysis. The Crimean observations were carried out in short time series by 4-6 consequent nights every month since September to March, in each season. The spectral regions including \Hal and \DNa lines were registered.

In one season,  2015-2016, multi-site monitoring of RY Tau and SU Aur was performed, including observations with the ALFOSC grism spectrograph at the 2.56-m Nordic Optical Telescope ($\lambda/\Delta\lambda = 15000$),  and the CAFE echelle spectrograph at the 2.2-m telescope at Centro Astronomico Hispano-Aleman (CAHA), $\lambda/\Delta\lambda=30000$. More information about the observations and data reductions was presented in our previous publication  \citepalias{Petrov2019}. 

%Several spectra were obtained in 2015-2017 with the Medium Resolution Spectrograph (MRES) at the 2.4-m Thai National Telescope (TNT) at the Thai National Observatory. 

As the data from our monitoring of these stars accumulated, the results obtained were published by \citet{Petrov2019,Petrov2021}. At present, our dataset contains 193 spectra (193 nights) of RY Tau and 176 spectra of SU Aur,  suitable for analysis of \Hal line variability. There are fewer spectra of the absorption \DNa lines, suitable for analysis,  since these lines are noisy on nights with poor image quality, and often contaminated with atmospheric water lines. On most of the nights of spectral observations, $BVR$ photometry of both targets was carried out at a 1.25 m telescope at CrAO.
The log of spectral observations and the V-magnitudes for the dates of observations are presented in Tables~\ref{tab:photRY} and \ref{tab:photSU} in the same format as in the  \citetalias{Petrov2019}.

In the next section, we analyze the \Hal line profile variability in the whole set of data obtained in 2013-2022:  9 seasons of RY Tau and 7 seasons of SU Aur.

\section{Results}
\label{sec:results}

\begin{figure}
	\center{\includegraphics[width=0.75\columnwidth]{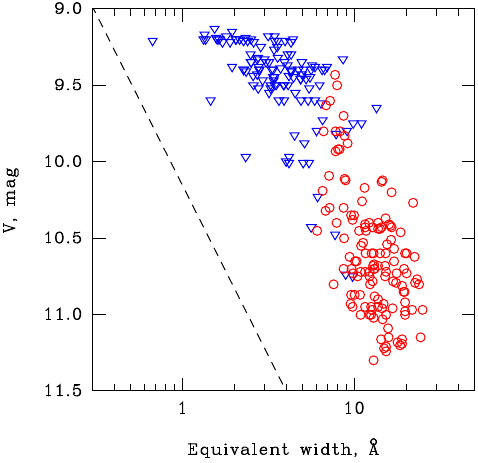}}
    \caption{Equivalent width of the \Hal emission as a function of stellar brightness in SU Aur (blue triangles) and RY Tau (red circles).}
    \label{fig:V_EWTOT}
\end{figure}

In both stars, RY Tau and SU Aur, the equivalent width of the \Hal emission increases as the star gets weaker (Fig.~\ref{fig:V_EWTOT}). This effect is caused by the circumstellar dust which obscures the star,  but not the whole region radiating in \Hal line.  The straight line in Fig.~\ref{fig:V_EWTOT} indicates the slope expected for the case of constant \Hal flux at variable stellar continuum. The observed scatter of \Hal equivalent widths is due to the intrinsic variability of the \Hal flux.

\begin{figure}
	\includegraphics[width=0.92\columnwidth]{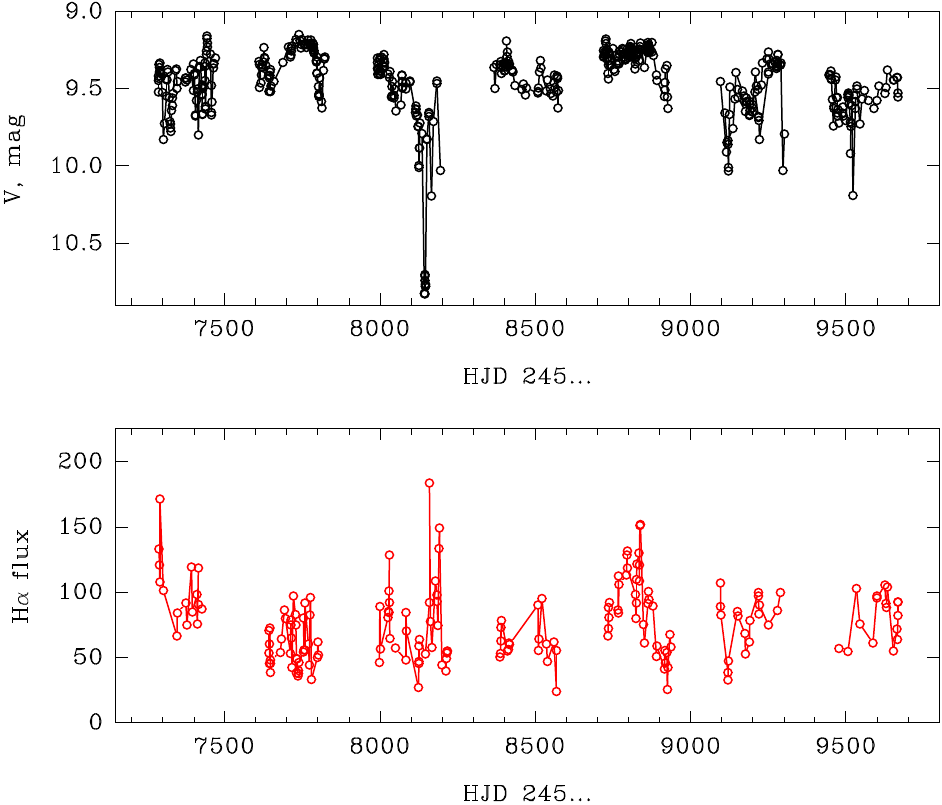}
    \caption{Light curve (above) and \Hal flux variations (below) of  SU Aur in 2015-2022. The \Hal flux is expressed in units  $3.67\times10^{-13}$ erg cm$^{-2}$ s$^{-1}$. }
    \label{fig:SU_V_Ha}
\end{figure}

\begin{figure}
	\includegraphics[width=0.95\columnwidth]{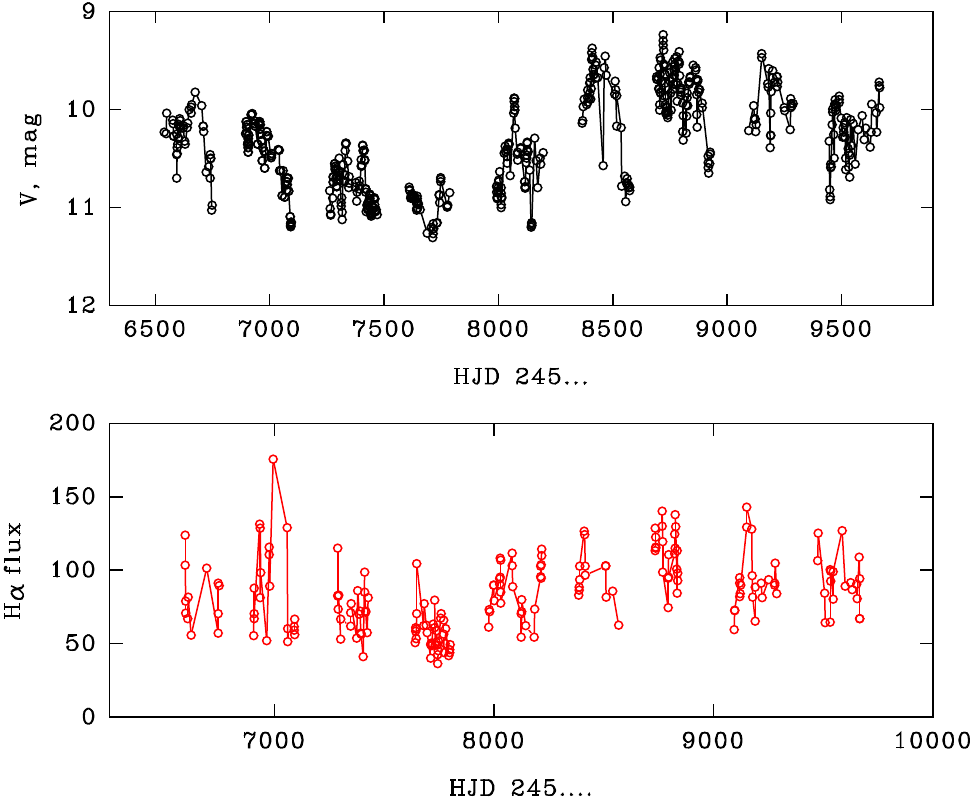}
    \caption{Light curve (above) and \Hal flux variations (below) of  RY Tau in 2013-2022. The \Hal flux units: $3.67\times10^{-13}$ erg cm$^{-2}$ s$^{-1}$.}
    \label{fig:RY_V_Ha}
\end{figure}

Figs.~\ref{fig:SU_V_Ha} and~\ref{fig:RY_V_Ha} show stellar brightness and \Hal flux variability in  SU Aur and RY Tau according to our observations,  presented in Tables~\ref{tab:photRY} and \ref{tab:photSU}. The \Hal flux is calculated as  F = EW$\times10^{-0.4(V-10)}$,  where the line flux is expressed in units of the continuum flux density of a star with $V$ = 10 mag,  which is $3.67\times10^{-13}$ erg cm$^{-2}$ s$^{-1}$. The magnitudes in the $R$ band would be more appropriate for this conversion, but only $V$ magnitudes are available to cover the full set of spectroscopic data. However we have checked that for RY Tau the color $V-R$ does not change considerably with brightness: on average, $V-R = 1.1 \pm 0.1$.  In the case of SU Aur the probable error in the \Hal flux, caused by variations in the $V-R$ color, is about 5\%, while  the \Hal flux varies by a factor of 3 and more (see Fig.~\ref{fig:SU_V_Ha}).

For analysis of spectral line profiles, the spectra obtained at different instruments were re-binned to get a common sampling of the wavelength scale. In the following analysis of \Hal line profile variations, we use the astrocentric radial velocity scale, assuming that the radial velocity of RY Tau and SU Aur is +18 \kms \citep{Petrov1996,Petrov1999}. Radial velocities variations due to possible photospheric inhomogeneities (spots) are expected to be less than a few \kms. A spot would also result in rotational modulation of stellar brightness. This effect was not found in RY Tau but may have been observed in SU Aur (see Table~\ref{tab:tab2}).

\subsection{\Hal profile variability in SU Aur}

\begin{figure}
\center{\includegraphics[width=0.8\columnwidth]{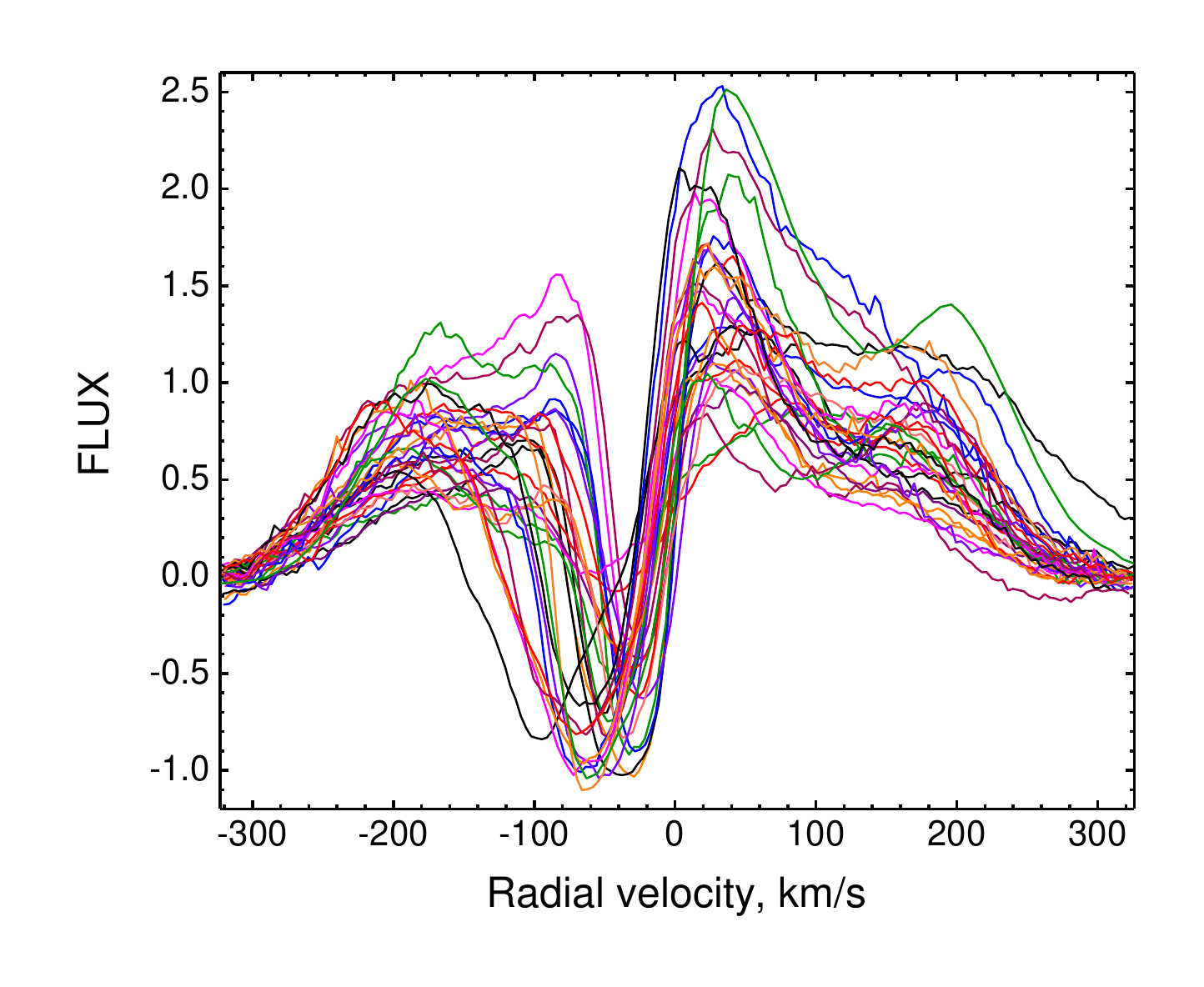}}
    \caption{A sample of \Hal line profile variability in SU Aur.}
    \label{fig:SU_Ha}
\end{figure}

A sample of 30 \Hal profiles in SU Aur, observed in different seasons of  2015-2022, is presented in Fig.~\ref{fig:SU_Ha} to show the pattern and amplitude of the line profile variability, typical for this star. All the 176 \Hal spectra were used  to quantify the line profile variability. 
 
\begin{figure}
\begin{minipage}{\linewidth}
\center{\includegraphics[width=\linewidth]{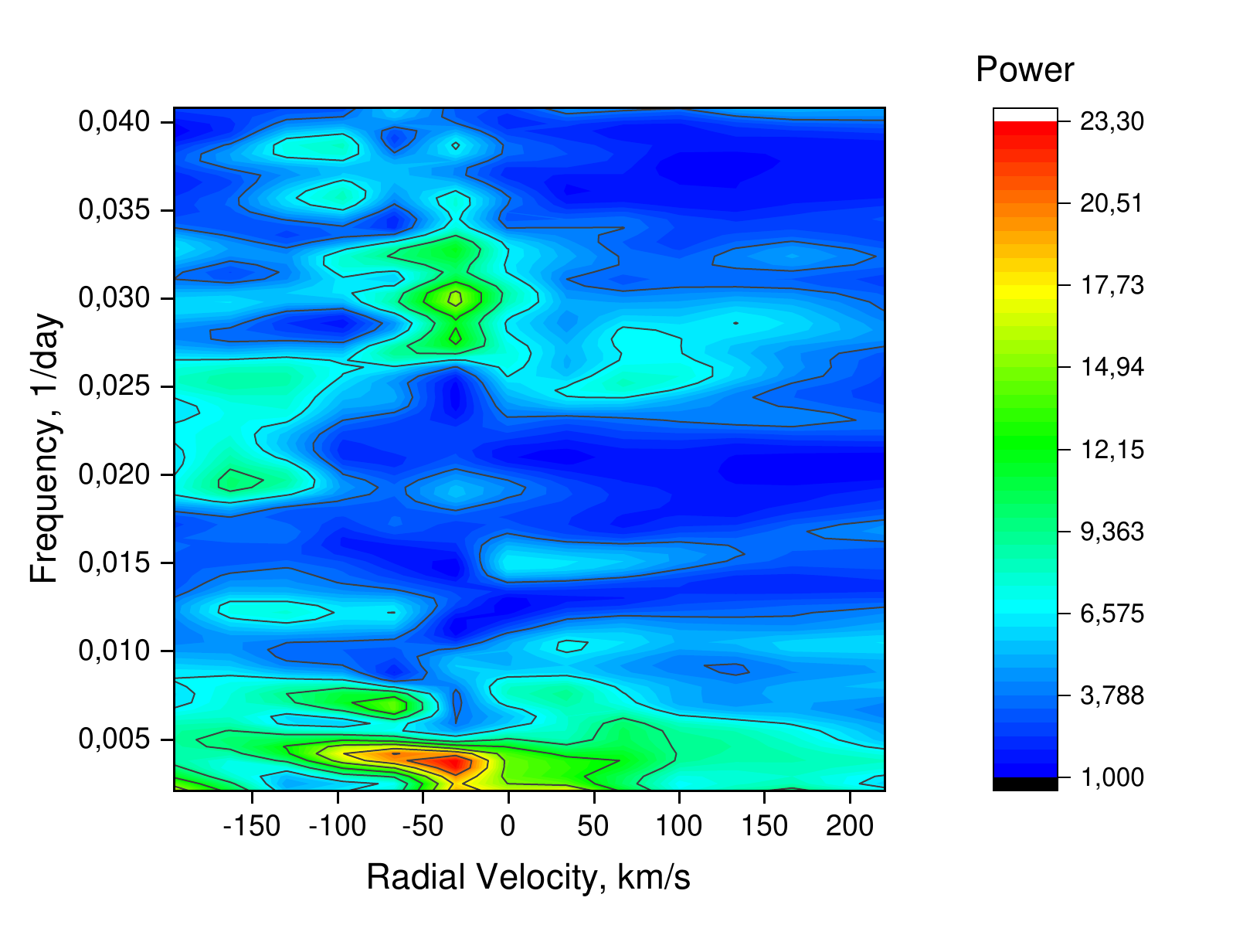}}
\end{minipage}
\vfill
\begin{minipage}{\linewidth}
\center{\includegraphics[width=0.7\linewidth]{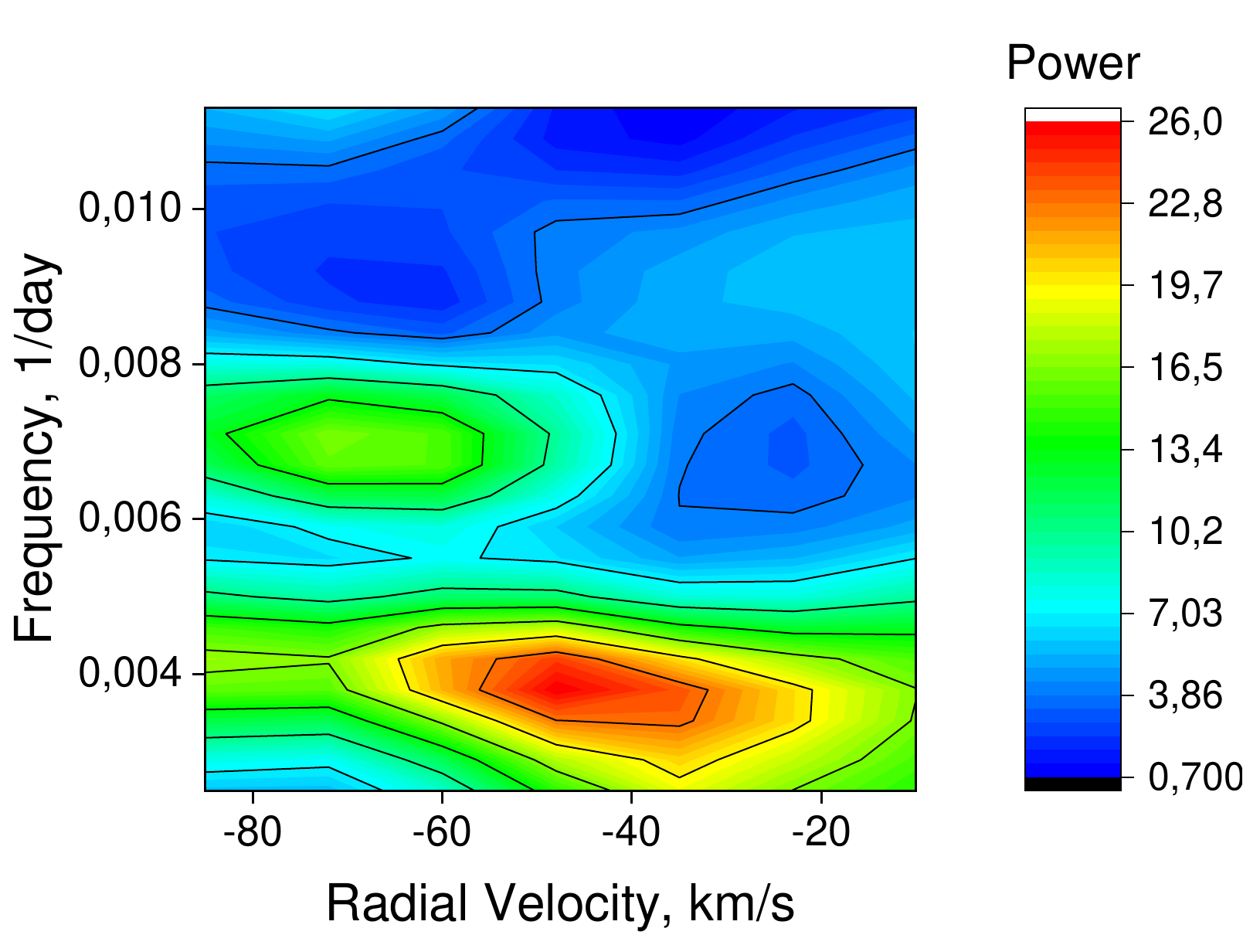}}
\end{minipage}
\caption{Upper: two-dimensional periodogram of the \Hal line flux in SU Aur. Lower: enlarged fragment of the periodogram. The red spot indicates the area of the most powerful oscillations with a period of $255\pm5$ days. }
\label{fig:SU_2d}
\end{figure}

%\begin{figure}
%\center{\includegraphics[width=\columnwidth]{SU_2D}}
 %   \caption{Two-dimensional periodogram of the \Hal line flux in SU Aur, (b) Fragment of the two-dimensional periodogram of the \Hal line flux in SU Aur. The red spot indicates the area of the most powerful oscillations with a period of 255 days.}
 %   \label{fig:SU_Ha}
%\end{figure}
A power spectrum of the flux variations was calculated in each radial velocity bin across the \Hal line profile. This results in a two-dimensional (2D) periodogram in the space ''Frequency vs Radial Velocity'', which reveals the areas of probable periodic signals. The power spectra were calculated following the procedure CORRELOGRAM described in \citet{Lawrence1987}. For better visibility of the area of a periodic signal, the power spectra were smoothed to a lower frequency resolution, about $10^{-3}$ day$^{-1}$.  The 2D periodogram of the \Hal line flux is shown in Fig.~\ref{fig:SU_2d}.  The power of the periodic oscillations is color-coded from blue to red as it increases. Most of the power is concentrated in the blueward of \Hal, at low frequencies. This region of the periodogram is expanded in the lower part of Fig.~\ref{fig:SU_2d}.  The most significant period P= $255\pm5$ days (1/P=0.004) is at a radial velocity RV = $-45 \pm 7$  \kms. The same result was obtained when only high-resolution (R $\geq27000$) spectra were used in the analysis.

\begin{figure}
\center{\includegraphics[width=0.93\columnwidth]{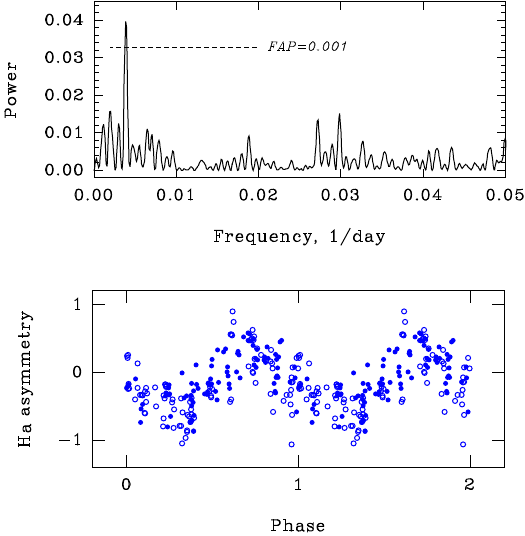}}
    \caption{Top panel: power spectrum of the \Hal profile asymmetry in SU Aur. Bottom panel: \Hal profile asymmetry, convolved with the period P = 255 days. Dots and circles correspond to the first and the second halves of the time series, correspondingly.}
    \label{fig:SU_pow}
\end{figure}

The period of 255 days is seen also in the ratio of fluxes measured at $-50$  and $+25$ \kms, corresponding to the deepest absorption and the highest emission in the profile.  This ratio may be considered as a parameter of the \Hal profile asymmetry.  The power spectrum and convolution of this parameter with a period of 255 days are presented in Fig.~\ref{fig:SU_pow}. The first and the second halves of the time series (2015-2018 and 2019-2022 respectively) are displayed with different symbols, to show that periodic variations persisted throughout the whole data set.  The same period is also present in the intensity variations of the blue-shifted absorptions in the \DNa lines.

\subsection{\Hal profile variability in RY Tau}

\begin{figure}
\center{\includegraphics[width=0.8\columnwidth]{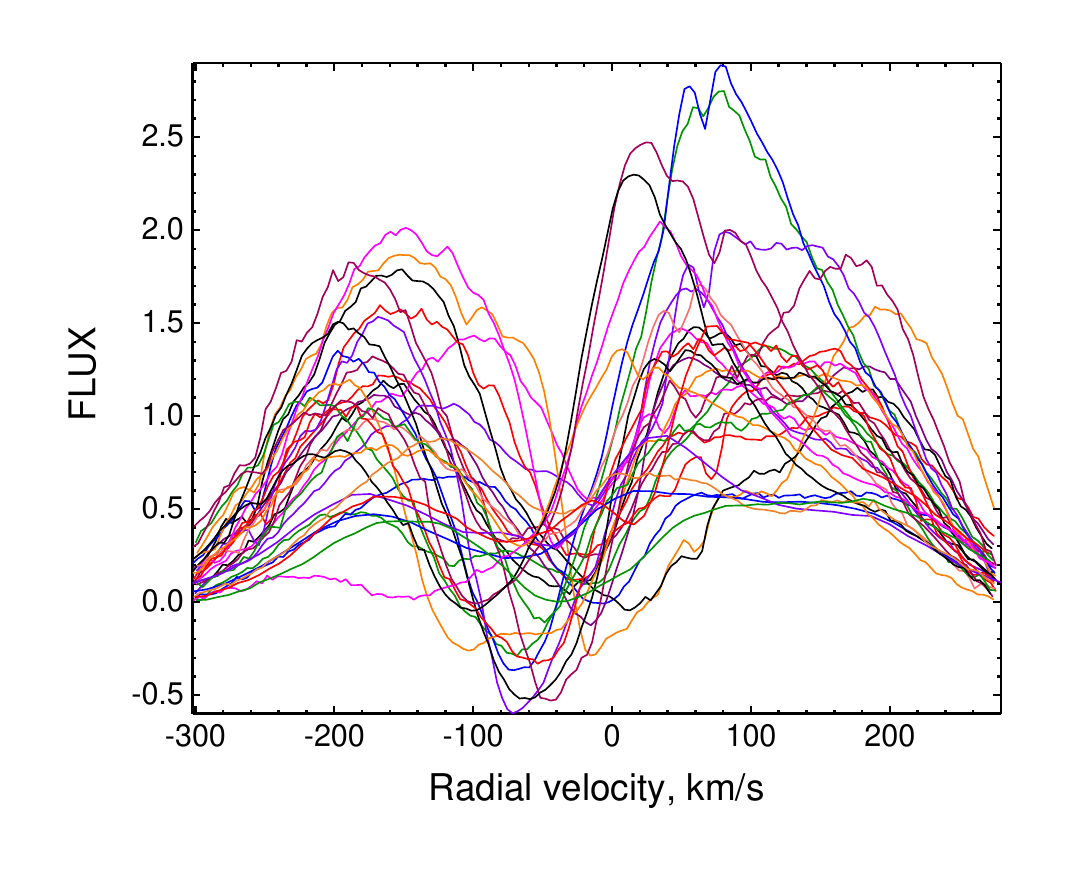}}
    \caption{A sample of \Hal line profile variability in RY Tau.}
    \label{fig:RY_Ha}
\end{figure}

A sample of  30  \Hal profiles in RY Tau, observed in different seasons between 2018 and 2021, is presented in Fig.~\ref{fig:RY_Ha}  showing the pattern and amplitude of the profile variability, typical for this star.  All the 193 \Hal spectra were used to quantify the line profile variability.

Periodic variations in the \Hal profile of RY Tau were previously reported in our publication, based on observations of  2013-2018: P=21.6 days at RV=-100 \kms \citep{Petrov2021}. A similar analysis has been done using the whole set of data (9 seasons), corresponding to the years 2013-2022.

\begin{figure}
\begin{minipage}{\linewidth}
\center{\includegraphics[width=\linewidth]{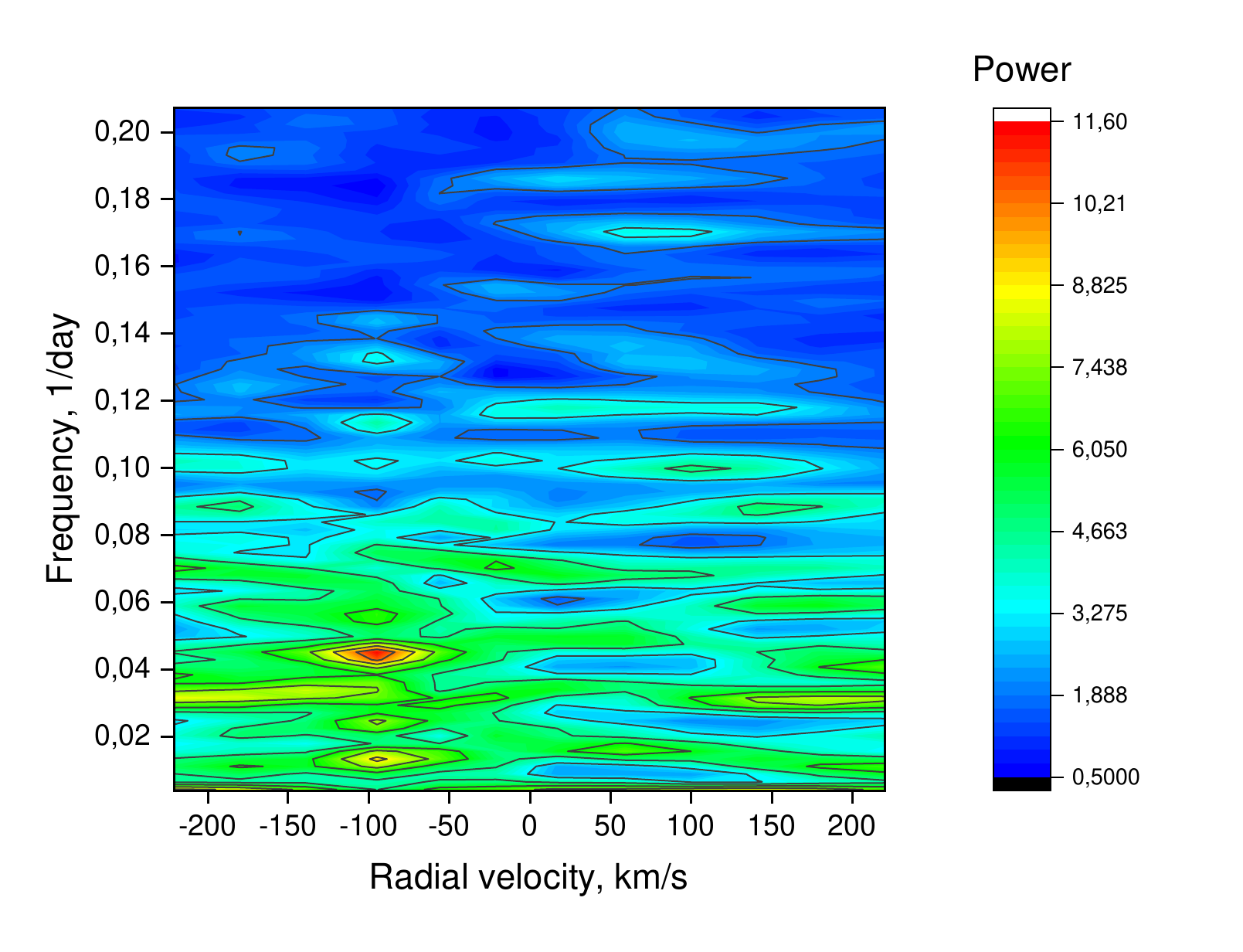}}
\end{minipage}
\vfill
\begin{minipage}{\linewidth}
\center{\includegraphics[width=0.7\linewidth]{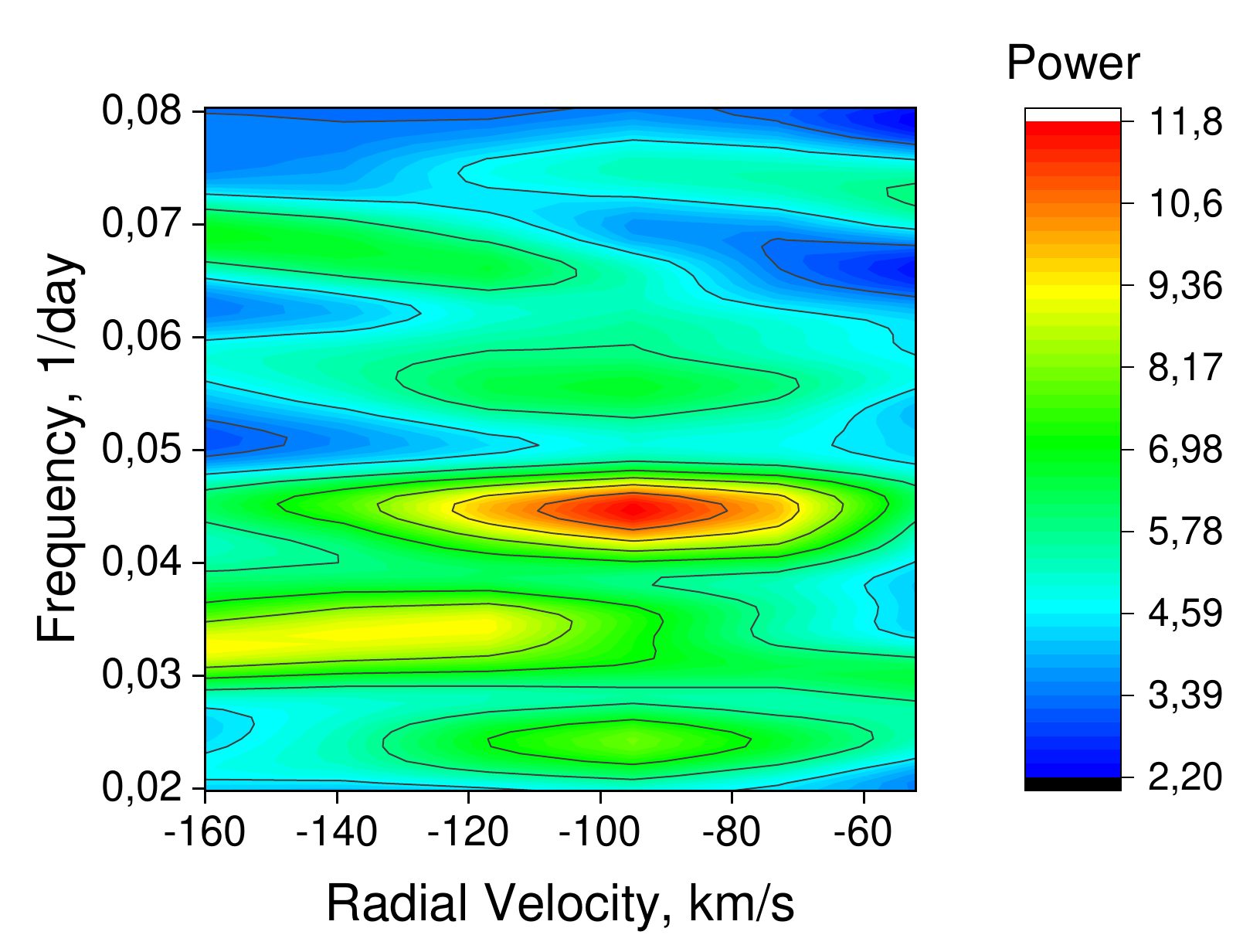}}
\end{minipage}
\caption{Upper: two-dimensional periodogram of the \Hal line flux in RY Tau. Lower: enlarged fragment of the periodogram. The red spot at RV=$-95\pm5$ \kms indicates the area of the most powerful oscillations with a period of 21.6 days.}
\label{fig:RY_2d}
\end{figure}

Fig.~\ref{fig:RY_2d} shows a 2D periodogram of the \Hal flux variations in RY Tau. The most significant period is present at radial velocity $-95\pm5$ \kms: the flux in this section of the profile varies with a period P= 21.6 days (1/P=0.0463). The lower part of the  Fig.~\ref{fig:RY_2d} shows this region with better resolution.

The period is also present in the variations of the line profile asymmetry. In the case of RY Tau, the \Hal line profile is less regular, as compared to SU Aur: the position of the blue-shifted  depression is not fixed at the same radial velocity, as well as the position of the red-shifted emission peak (see Fig.~\ref{fig:RY_Ha}). Therefore, to measure the line asymmetry we use a more integrated parameter:  (b-r)/(b+r), where 'b' and 'r' are fluxes in the 'blue' and 'red' halves of \Hal profile, relative to zero radial velocity. The most negative values of the asymmetry correspond to the P Cyg type profile.

\begin{figure}
\center{\includegraphics[width=0.89\columnwidth]{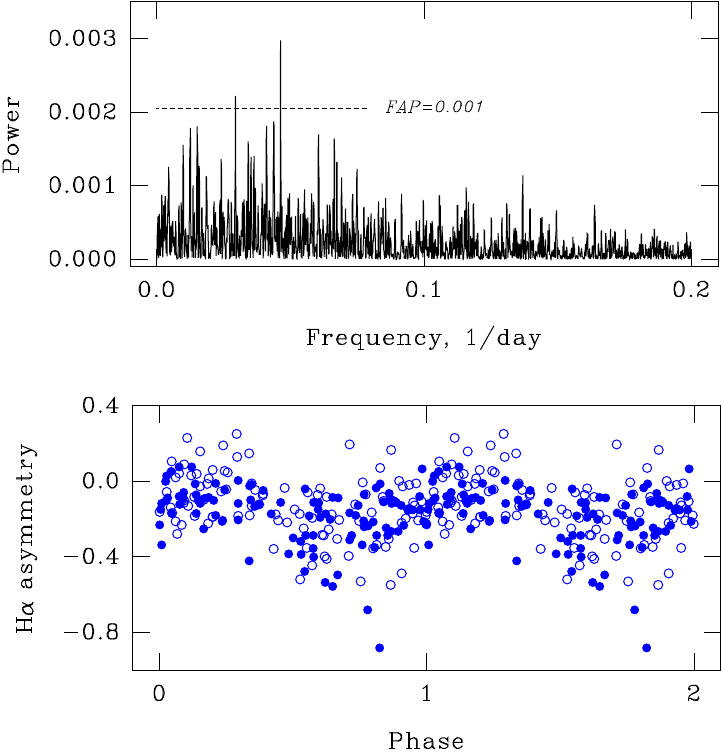}}
    \caption{Top panel: Power spectrum of \Hal profile asymmetry in RY Tau. Bottom panel: \Hal profile asymmetry convolved with the period of 21.6 days. Dots and circles correspond to the first and the second halves of the time series, correspondingly.}
    \label{fig:RY_pow}
\end{figure}

The power spectrum of \Hal line asymmetry and convolution of the asymmetry with the period of 21.6 days, are shown in Fig.~\ref{fig:RY_pow}.  About the same sinusoidal signal with the period of 21.6 days appears when the asymmetry of \Hal is expressed as the ratio of flux at radial velocity -95 \kms to the total flux in the line:   F$_{-95}/$F$_{total}$.  The period of 21.6 days is also detected in the variability of the blue-shifted absorption of the \DNa lines, at a radial velocity of about -100 \kms \citep{Petrov2021}.
%The same amplitude of sinusoidal signal is revealed when the asymmetry of \Hal is expressed as the ratio of flux at radial velocity -95 \kms to the total flux in the line: F$_{-95}/$F$_{total}$.

The level of false alarm probability FAP=0.001, indicated in Figs.~\ref{fig:SU_pow} and~\ref{fig:RY_pow}, was estimated via the Fisher's Method of Randomization \citep{Nm1985}. The highest peaks in the power spectra of SU Aur (Fig.~\ref{fig:SU_pow}) and RY Tau (Fig.~\ref{fig:RY_pow}) correspond to the ''red spots'' in the 2D periodograms in Figs.~\ref{fig:SU_2d} and \ref{fig:RY_2d}. In the case of RY Tau, the power spectrum shows two peaks above the FAP=0.001. The less significant peak indicates variations with a period of about 34 days in the blue wing of \Hal emission. In the 2D periodogram (Fig.~\ref{fig:RY_2d}) it looks like a yellow strip at frequency 1/P $\sim$ 0.03, starting from RV= -110 \kms.

\section{Discussion}
\label{sec:discussion}

The light variations of SU Aur and RY Tau, observed during our spectroscopic monitoring (Figs.~\ref{fig:SU_V_Ha} and~\ref{fig:RY_V_Ha}),  are typical for these stars. Taking into account the large inclination of the accretion discs, the short-term dimmings of SU Aur may be explained as a variable extinction in the dusty disc wind \citep{Labdon2019}. In RY Tau, the brightness variations over one magnitude can also be explained as due to circumstellar dust, because the veiling in the visible region of the photospheric spectrum remains low, < 0.1 \citep{Petrov1999}. 

In both stars, RY Tau and SU Aur, the periodic changes appear in the blue wings of \Hal and the \DNa lines, in the radial velocities range that correspond to absorption in the wind. In the case of RY Tau, the period (21.6 days) could be assigned to the quasi-periodic outflows in the regime of the magnetic propeller. Therefore, in our previous publication \citep{Petrov2021} we discussed two scenarios: 1) a stable magnetic propeller, or 2) a Keplerian motion of a planet at 0.2 AU. In the case of SU Aur, the period is much longer and cannot be identified with the magnetic propeller regime. Therefore, we have to consider whether the periods are related to rotation in the accretion disk.

Accretion discs around CTTSs are expected to have magnetic fields, driving the disc wind out of the disc surface. The poloidal magnetic field lines diverge outward due to the inward flow of matter in the disc plane and are twisted by the disc rotation. Thus, the ionized gas from the disc surface is accelerated into the wind by the magneto-centrifugal forces \citep{Blandford1982}.  Azimuthal asymmetries in the density distribution of the accretion disc lead to asymmetries in the wind.

\begin{figure}
\center{\includegraphics[width=0.8\columnwidth]{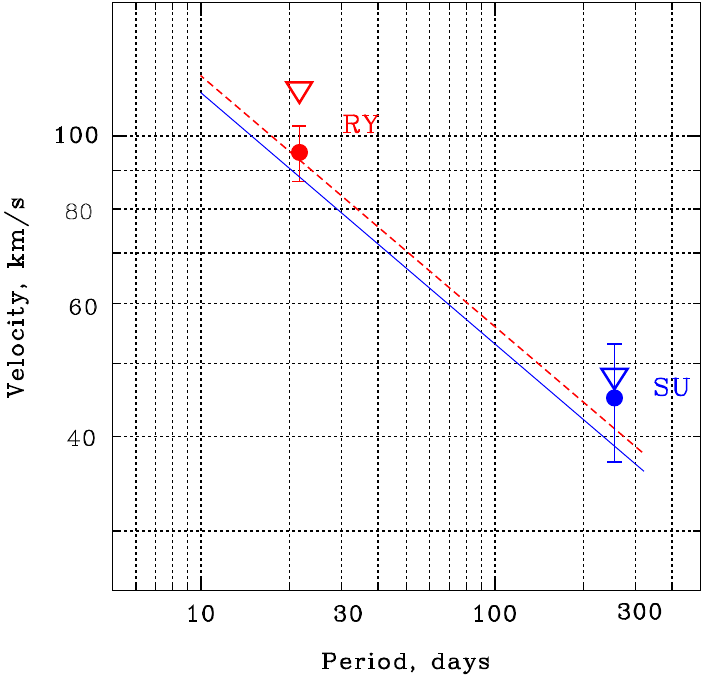}}
\caption{Two straight lines show the Keplerian relationship between the orbital velocity and the orbital period for the masses of RY Tau (red dashed line) and SU Aur (blue solid line).  Dots with error bars show the observed radial velocities and periods in the \Hal flux variations in RY Tau and SU Aur. Open triangles indicate the deprojected radial velocities.}
\label{fig:KEPLER}
\end{figure}

The masses of RY Tau and SU Aur are known, so we can check whether the periods found  are related to the Keplerian motion. In Fig.~\ref{fig:KEPLER} the straight lines show the Keplerian relationship between the orbital velocity and the orbital period for the masses of RY Tau and SU Aur. The dots with error bars indicate the observed values of the periodic variations of the \Hal flux,  and the radial velocities, at which these changes occur. 

%This diagram shows that the observed periods and velocities do not contradict the hypothesis of the Keplerian rotation of some structure at the base of the disc wind.

The spatial velocity of a gas particle in a disk wind may be considered as the sum of two vectors: toroidal and poloidal \citep[e.g.][]{Kurosawa2006}. A gas partical starts from the surface of the disk and moves along an expanding spiral, rising above the disk plane. The blue-shifted absorption in the Ha emission line profile is formed in the wind along the line of sight to the star. The observed radial velocity of this absorption is defined mainly by the wind acceleration in poloidal direction. The projection of the poloidal component to the line of sight depends on two angles: the inclination of the disc axis $i$, and the inclination of the wind stream to the disc axis (the semi-open angle), which must be $\geq 30^\circ$ in order to start the centrifugal 
acceleration \citep{Blandford1982}. Hence, the deprojected poloidal velocity: $V_p \leq RV/\cos(i-30)$. In  Fig.~\ref{fig:KEPLER}, the estimated poloidal velocities of wind are shown with triangles.

In the magneto-centrifugal wind model, the terminal velocity of wind is proportional to the Keplerian velocity at the launching point on the accretion disc \citep{Ferreira2006}. In the case of high inclination $i$, the line of sight to the star crosses the wind streams launched from different radii in the disc plane. Assuming there is an axial asymmetry in the wind, caused by an azimuthal asymmetry in the structure of the accretion disc at a certain radius R$_w$, the column density of the disc wind on the line of sight to the star varies with a Keplerian period, corresponding to the orbital radius R$_w$.

The observed periodic changes in the disc wind do not yet mean the presence of a planet in the disc. Moreover, a planet by itself does not produce a density enhancement in the disc wind. There must be some feature in the accretion disc that rotates at Keplerian velocity and can create a non-axisymmetric dense structure in the wind. 

What causes this asymmetry? Images of accretion discs of young stars obtained with the ALMA interferometer show that the surface brightness of the discs is significantly inhomogeneous, with different substructures, presumably reflecting different instabilities in the disc leading to maxima of gas pressure, or indicating the locations of planet formation  \citep{Andrews2020}. 

Possible candidates for azimuthal asymmetry in the disc are vortices formed due to the Rossby wave instability \citep{Lovelace1999,Li2000}. They typically form in parts of the disc with a sharp density gradient, such as at the dead zone boundaries, dust sublimation radii, magnetospheric cavity boundary, or the edges of gaps opened by giant planets \citep[e.g.][]{Armitage2019}. A vortex represents the region of high pressure and density, which attracts dust particles. The vortex moves around a star approximately with a Keplerian velocity and rotates around the center of the vortex.

The derived orbital radius of 0.2 AU in RY Tau is very close to the inner dusty disc radius of $0.2-0.3$ AU measured from near IR interferometry \citep[e.g.][]{Perraut2021} where some density jump is expected and vortices may form.  Even higher density jump is expected at the inner dead zone transition \citep{Gammie1996}, where the inner parts of the disc are strongly ionized and matter accretes more rapidly due to the MRI-driven turbulence \citep[e.g.][]{Balbus1991}, while accretion rate is low in the external, weakly-ionized parts of the disc.  This density jump is shown to be a region where vortexes form \citep[e.g.][]{Varniere2006,Lyra2012}.  The derived radii of $0.2-0.9$ AU for the location of the density enhancements in the disc are in the range of dead zones expected in considered stars. Simulations show that typically, the vortex is stable during many rotations  \citep[e.g.][]{Li2001}, however detailed radiative simulations by \citet{Faure2015} show that at the dead zone boundaries vortexes form episodically and migrate inward in a cyclic manner. On average, the vortex position coincides with the location of the dead zone, and the period of azimuthal rotation will change if the dead zone will change its position. 

If the vortex forms at the edge of the gap formed by a giant planet, then its period will be associated with a period of the planet and is expected to be more stable for a long time because a planet migrates slowly \citep[see review by][]{Armitage2019}. Periods obtained in our work can be connected with one of the above mechanisms.

Outflows from Rossby vortices are expected to have a higher density than those from surrounding parts of the disc and may produce a local stream of denser matter in the wind. The poloidal magnetic field threading the disc does not prevent the formation of Rossby vortices and driving matter into the wind \citep{Yu2013}. Moreover, the rotation of the vortex leads to the generation of an additional, azimuthal magnetic field around the vortex which may contribute to driving denser matter into the wind \citep{Lovelace2014,Matilsky2018}.

%%%[то, что мелким шрифтом – закомментировать. Возможно, потом понадобится]
%%%In the disc wind model, the wind starts from the disc surface with an initial velocity, scaled as the local Keplerian velocity. The wind is then accelerated by the magnetic centrifuge of the rotating disc. As a result, a wind particle moves along an expanding spiral, raising above the disc surface, as shown schematically on Fig…_Assuming that a denser wind starts from some area on the disc surface, one can expect the formation of a spiral of a denser flow. For an observer, it looks like a ’cloud’ of dense wind that regularly crosses the line of sight to the star. 

%The disc wind geometry can be represented as the sum of two velocity components: azimuthal (orbital) and poloidal  \citep[e.g.][]{Kurosawa2006,Tambovtseva2014}.  In this model, the wind starts from the disc surface with an initial velocity, scaled as the local Keplerian velocity.  The wind is then accelerated by the magnetic centrifuge of the rotating disc. As a result, a wind particle moves along an expanding spiral, raising above the disc surface. Assuming that a denser wind starts from some area on the disc surface, one can expect the formation of a spiral of a denser flow. For an observer, it looks like a 'cloud' of dense wind that regularly crosses the line of sight to the star.  The observed radial velocity of the wind stream depends on the inclination of the disc, being more negative at a larger inclination \citep{Kurosawa2011}. 

The key condition for the stability of a stream of dense wind is the prevalence of magnetic pressure in the disc wind. There is the Alfven surface below which the magnetic pressure exceeds the gas pressure and the kinetic pressure: B$^2/8\pi  > $ P$_g$  + $\rho \cdot$ v$^2$.  Therefore, the streams of higher density in the wind cannot dissipate until they go above the Alfven surface. Below the Alfven surface, the disc wind  'remembers' the structure of the disc surface. This is a kind of magnetic 'memory' of the disc wind: it is permanently written at the base of the wind and erased above the Alfven surface.

These results open up new perspectives in studying the structure of the inner regions of the accretion discs. A method for mapping structural features on a moving surface, using changes in spectral line profiles, is known as the Doppler Imaging technique. It is widely applied in mapping spots on stars, where the inclination of the rotational axis and the period of rotation are well known. A similar technique, Doppler tomography, is applied for the analysis of the accretion disc emission line regions in cataclysmic variables  \citep{Marsh1988}.  

In the case of Classical T Tauri stars, by studying the density streams in the disc wind, we learn about the inner accretion disc structure and the orbital distances, related to planets formation.  The method may be applied for the discs with high inclination, where the line of sight crosses the disc wind.

\section{Conclusions}

Spectroscopic monitoring of CTTS provides information about the gas flows around the stars. In the case of high inclination of the accretion disc, when the line of sight crosses the region of the disc wind below Alfven's surface, observations may reveal azimuthal asymmetry in the disc wind's density caused by the structural features of the disc surface.  Our time series of RY Tau and  SU Aur revealed stable periodic variations in the strength of the blue-shifted absorptions in \Hal and \DNa lines. The radial velocity of these variable absorptions, the periods of variations, and the masses of the stars are found to be related by Kepler's law.  The structural details in the discs, presumably associated with Rossby waves, appeared at a distance of 0.2 AU in RY Tau and 0.9 AU in SU Aur.  These findings open a possibility to research the structure of the inner accretion disc of a CTTS on scales of the order of one AU or less and to reveal the orbital distances, related to the formation of planets.

\section*{Acknowledgements}

Partly based on observations made with the Nordic Optical Telescope, owned in collaboration by the University of Turku and Aarhus University, and operated jointly by Aarhus University, the University of Turku, and the University of Oslo, representing Denmark, Finland, and Norway, the University of Iceland and Stockholm University at the Observatorio del Roque de los Muchachos, La Palma, Spain, of the Instituto de Astrofisica de Canarias, using ALFOSC, which is provided by the Instituto de Astrofisica de Andalucia (IAA) under a joint agreement with the University of Copenhagen and NOT.

The Crimean observations in 2019-2020 by  S. Artemenko, E. Babina, and P. Petrov were supported by a grant from the Russian Science Foundation 19-72-10063. The work of  S. Yu. Gorda was partially supported by the Ministry of Science and Higher Education of Russia, topic No. FEUZ-2023-0019. M. M. Romanova was supported in part by NSF grant AST-2009820. J. F. Gameiro was supported by funda\c{c}\~{a}o para a Ci\^{e}ncia e Tecnologia (FCT) through the research grants UIDB/04434/2020 and UIDP/04434/2020.

We thank R. V. E. Lovelace for useful discussions and an anonymous referee for a thorough review that improved the paper.

\section*{Data availability statement}

The data underlying this article will be shared on reasonable request to the corresponding author [PP].



\bibliographystyle{mnras}
\bibliography{references} % if your bibtex file is called example.bib

\begin{thebibliography}{}
\makeatletter
\relax
\def\mn@urlcharsother{\let\do\@makeother \do\$\do\&\do\#\do\^\do\_\do\%\do\~}
\def\mn@doi{\begingroup\mn@urlcharsother \@ifnextchar [ {\mn@doi@}
  {\mn@doi@[]}}
\def\mn@doi@[#1]#2{\def\@tempa{#1}\ifx\@tempa\@empty \href
  {http://dx.doi.org/#2} {doi:#2}\else \href {http://dx.doi.org/#2} {#1}\fi
  \endgroup}
\def\mn@eprint#1#2{\mn@eprint@#1:#2::\@nil}
\def\mn@eprint@arXiv#1{\href {http://arxiv.org/abs/#1} {{\tt arXiv:#1}}}
\def\mn@eprint@dblp#1{\href {http://dblp.uni-trier.de/rec/bibtex/#1.xml}
  {dblp:#1}}
\def\mn@eprint@#1:#2:#3:#4\@nil{\def\@tempa {#1}\def\@tempb {#2}\def\@tempc
  {#3}\ifx \@tempc \@empty \let \@tempc \@tempb \let \@tempb \@tempa \fi \ifx
  \@tempb \@empty \def\@tempb {arXiv}\fi \@ifundefined
  {mn@eprint@\@tempb}{\@tempb:\@tempc}{\expandafter \expandafter \csname
  mn@eprint@\@tempb\endcsname \expandafter{\@tempc}}}

\bibitem[\protect\citeauthoryear{Agra-Amboage, Dougados, Cabrit, Garcia  \&
  Ferruit}{Agra-Amboage et~al.}{2009}]{Agra2009}
Agra-Amboage V.,  Dougados C.,  Cabrit S.,  Garcia P. J.~V.,   Ferruit P.,
  2009, A\&A, 493, 1029

\bibitem[\protect\citeauthoryear{Akiyama, Vorobyov, Liu, Dong, de Leon, Liu  \&
  Tamura}{Akiyama et~al.}{2019}]{Akiyama2019}
Akiyama E.,  Vorobyov E.~I.,  Liu H.~B.,  Dong R.,  de Leon J.,  Liu S.-Y.,
  Tamura M.,  2019, AJ, 157, 165

\bibitem[\protect\citeauthoryear{Alencar et~al.,}{Alencar
  et~al.}{2018}]{Alencar2018}
Alencar S. H.~P.,  et~al., 2018, A\&A, 620, 195

\bibitem[\protect\citeauthoryear{Andrews}{Andrews}{2020}]{Andrews2020}
Andrews S.~M.,  2020, ARA\&A, 58, 483

\bibitem[\protect\citeauthoryear{Armitage \& Kley}{Armitage \&
  Kley}{2019}]{Armitage2019}
Armitage P.~J.,  Kley W.,  2019, From Protoplanetary Disks to Planet Formation,
  Saas-Fee Advanced Course, Volume 45. Swiss Society for Astrophysics and
  Astronomy.
Berlin: Springer Berlin Heidelberg

\bibitem[\protect\citeauthoryear{Artemenko, Grankin  \& Petrov}{Artemenko
  et~al.}{2012}]{Artemenko2012}
Artemenko S.~A.,  Grankin K.~N.,   Petrov P.~P.,  2012, AstL, 38, 783

\bibitem[\protect\citeauthoryear{Balbus \& Hawley}{Balbus \&
  Hawley}{1991}]{Balbus1991}
Balbus S.~A.,  Hawley J.~F.,  1991, ApJ, 376, 214

\bibitem[\protect\citeauthoryear{Blandford \& Payne}{Blandford \&
  Payne}{1982}]{Blandford1982}
Blandford R.~D.,  Payne D.~G.,  1982, MNRAS, 199, 883

\bibitem[\protect\citeauthoryear{Bouvier}{Bouvier}{1990}]{Bouvier1990}
Bouvier J.,  1990, AJ, 99, 946

\bibitem[\protect\citeauthoryear{Bouvier, Alencar, Harries, Johns-Krull  \&
  Romanova}{Bouvier et~al.}{2007a}]{Bouvier2007b}
Bouvier J.,  Alencar S.,  Harries T.,  Johns-Krull C.~M.,   Romanova M.~M.,
  2007a, in Protostars and Planets V. University of Arizona Press, pp 479--494

\bibitem[\protect\citeauthoryear{Bouvier, Alencar, Boutelier, Dougados, Balog,
  Grankin, Hodgkin  \& et al.}{Bouvier et~al.}{2007b}]{Bouvier2007a}
Bouvier J.,  Alencar S. H.~P.,  Boutelier T.,  Dougados C.,  Balog Z.,  Grankin
  K.,  Hodgkin S.~T.,   et al. 2007b, A\&A, 463, 1017

\bibitem[\protect\citeauthoryear{Bouvier et~al.,}{Bouvier
  et~al.}{2023}]{Bouvier2023}
Bouvier J.,  et~al., 2023, A\&A, 672, 5

\bibitem[\protect\citeauthoryear{Calvet, Muzerolle, Briceno, Hernandez,
  Hartmann, Saucedo  \& Gordon}{Calvet et~al.}{2004}]{Calvet2004}
Calvet N.,  Muzerolle J.,  Briceno C.,  Hernandez J.,  Hartmann L.,  Saucedo
  J.~L.,   Gordon K.~D.,  2004, AJ, 128, 1294

\bibitem[\protect\citeauthoryear{Cody, Tayar, Hillenbrand, Matthews  \&
  Kallinger}{Cody et~al.}{2013}]{Cody2013}
Cody A.~M.,  Tayar J.,  Hillenbrand L.~A.,  Matthews J.~M.,   Kallinger T.,
  2013, AJ, 145, 79

\bibitem[\protect\citeauthoryear{Davies et~al.,}{Davies
  et~al.}{2020}]{Davies2020}
Davies C.~L.,  et~al., 2020, ApJ, 897, 31

\bibitem[\protect\citeauthoryear{DeWarf, Sepinsky, Guinan, Ribas  \&
  Nadalin}{DeWarf et~al.}{2003}]{DeWarf2003}
DeWarf L.~E.,  Sepinsky J.~F.,  Guinan E.~F.,  Ribas I.,   Nadalin I.,  2003,
  ApJ, 590, 357

\bibitem[\protect\citeauthoryear{Faure, Fromang, Latter  \& Meheut}{Faure
  et~al.}{2015}]{Faure2015}
Faure J.,  Fromang S.,  Latter H.,   Meheut H.,  2015, A\&A, 573, 132

\bibitem[\protect\citeauthoryear{Ferreira, Dougados  \& Cabrit}{Ferreira
  et~al.}{2006}]{Ferreira2006}
Ferreira J.,  Dougados C.,   Cabrit S.,  2006, A\&A, 453, 785

\bibitem[\protect\citeauthoryear{Gahm, Gullbring, Fischerstrom, Lindroos  \&
  Loden}{Gahm et~al.}{1993}]{Gahm1993}
Gahm G.~F.,  Gullbring E.,  Fischerstrom C.,  Lindroos K.~P.,   Loden K.,
  1993, A\&AS, 100, 371

\bibitem[\protect\citeauthoryear{Gammie}{Gammie}{1996}]{Gammie1996}
Gammie C.,  1996, ApJ, 457, 355

\bibitem[\protect\citeauthoryear{Garufi, Podio, Bacciotti, Antoniucci  \& et
  al.}{Garufi et~al.}{2019}]{Garufi2019}
Garufi A.,  Podio L.,  Bacciotti F.,  Antoniucci S.,   et al. 2019, A\&A, 628,
  68

\bibitem[\protect\citeauthoryear{Giampapa, Basri, Johns  \& Imhoff}{Giampapa
  et~al.}{1993}]{Giampapa1993}
Giampapa M.~S.,  Basri G.~S.,  Johns C.~M.,   Imhoff C.,  1993, ApJS, 89, 321

\bibitem[\protect\citeauthoryear{Ginski et~al.,}{Ginski
  et~al.}{2021}]{Ginski2021}
Ginski C.,  et~al., 2021, ApJ, 908, 25

\bibitem[\protect\citeauthoryear{Grandjean et~al.,}{Grandjean
  et~al.}{2021}]{Grandjean2021}
Grandjean A.,  et~al., 2021, A\&A, 650, 39

\bibitem[\protect\citeauthoryear{Grankin, Melnikov, Bouvier, Herbst  \&
  Shevchenko}{Grankin et~al.}{2007}]{Grankin2007}
Grankin K.~N.,  Melnikov S.~Y.,  Bouvier J.,  Herbst W.,   Shevchenko V.~S.,
  2007, A\&A, 461, 183

\bibitem[\protect\citeauthoryear{Herbig, Petrov  \& Duemmler}{Herbig
  et~al.}{2003}]{Herbig2003}
Herbig G.~H.,  Petrov P.~P.,   Duemmler R.,  2003, ApJ, 595, 384

\bibitem[\protect\citeauthoryear{Herbst \& Stine}{Herbst \&
  Stine}{1984}]{Herbst1984}
Herbst W.,  Stine P.~C.,  1984, AJ, 89, 1716

\bibitem[\protect\citeauthoryear{Herbst et~al.,}{Herbst
  et~al.}{1987}]{Herbst1987}
Herbst W.,  et~al., 1987, AJ, 94, 137

\bibitem[\protect\citeauthoryear{Herbst, Herbst, Grossman  \& Weinstein}{Herbst
  et~al.}{1994}]{Herbst1994}
Herbst W.,  Herbst D.~K.,  Grossman E.~J.,   Weinstein D.,  1994, AJ, 108, 1906

\bibitem[\protect\citeauthoryear{Isella, Carpenter  \& Sargent}{Isella
  et~al.}{2010}]{Isella2010}
Isella A.,  Carpenter J.~M.,   Sargent A.~I.,  2010, ApJ, 714, 1746

\bibitem[\protect\citeauthoryear{Ismailov, Adigezalzade  \&
  Bahaddinova}{Ismailov et~al.}{2015}]{Ismailov2015}
Ismailov N.~Z.,  Adigezalzade A.~N.,   Bahaddinova G.~R.,  2015, PKAS, 30, 229

\bibitem[\protect\citeauthoryear{Johns \& Basri}{Johns \&
  Basri}{1995}]{Johns1995}
Johns C.,  Basri G.,  1995, AJ, 109, 2800

\bibitem[\protect\citeauthoryear{Kurosawa \& Romanova}{Kurosawa \&
  Romanova}{2012}]{Kurosawa2012}
Kurosawa R.,  Romanova M.~M.,  2012, MNRAS, 426, 2901

\bibitem[\protect\citeauthoryear{Kurosawa, Harries  \& Symington}{Kurosawa
  et~al.}{2006}]{Kurosawa2006}
Kurosawa R.,  Harries T.~J.,   Symington N.~H.,  2006, MNRAS, 370, 580

\bibitem[\protect\citeauthoryear{Kurosawa, Romanova  \& Harries}{Kurosawa
  et~al.}{2011}]{Kurosawa2011}
Kurosawa R.,  Romanova M.~M.,   Harries T.~J.,  2011, MNRAS, 416, 2623

\bibitem[\protect\citeauthoryear{Labdon et~al.,}{Labdon
  et~al.}{2019}]{Labdon2019}
Labdon A.,  et~al., 2019, A\&A, 561, 23

\bibitem[\protect\citeauthoryear{Li, Finn, Lovelace  \& A.}{Li
  et~al.}{2000}]{Li2000}
Li H.,  Finn J.~M.,  Lovelace R. V.~E.,   A. C.~S.,  2000, ApJ, 533, 1023

\bibitem[\protect\citeauthoryear{Li, Colgate, Wendroff  \& Liska}{Li
  et~al.}{2001}]{Li2001}
Li H.,  Colgate S.~A.,  Wendroff B.,   Liska R.,  2001, ApJ, 551, 874

\bibitem[\protect\citeauthoryear{Long et~al.,}{Long et~al.}{2019}]{Long2019}
Long F.,  et~al., 2019, ApJ, 882, 49

\bibitem[\protect\citeauthoryear{Lovelace \& Romanova}{Lovelace \&
  Romanova}{2014}]{Lovelace2014}
Lovelace R. V.~E.,  Romanova M.~M.,  2014, FlDyR, 46, 1401

\bibitem[\protect\citeauthoryear{Lovelace, Li, Colgate  \& Nelson}{Lovelace
  et~al.}{1999}]{Lovelace1999}
Lovelace R. V.~E.,  Li H.,  Colgate S.~A.,   Nelson A.~F.,  1999, ApJ, 513, 805

\bibitem[\protect\citeauthoryear{Lyra \& Mac~Low}{Lyra \&
  Mac~Low}{2012}]{Lyra2012}
Lyra W.,  Mac~Low M.-M.,  2012, ApJ, 756, 62

\bibitem[\protect\citeauthoryear{Marsh \& Horne}{Marsh \&
  Horne}{1988}]{Marsh1988}
Marsh T.,  Horne K.,  1988, MNRAS, 235, 269

\bibitem[\protect\citeauthoryear{Matilsky, Dyda, Lovelace  \& Lii}{Matilsky
  et~al.}{2018}]{Matilsky2018}
Matilsky L.,  Dyda S.,  Lovelace R. V.~E.,   Lii P.~S.,  2018, MNRAS, 480, 3671

\bibitem[\protect\citeauthoryear{Nemec \& Nemec}{Nemec \& Nemec}{1985}]{Nm1985}
Nemec A. F.~L.,  Nemec J. M.~A.,  1985, AJ, 90, 2317

\bibitem[\protect\citeauthoryear{Nguyen, Brandeker, van Kerkwijk  \&
  Jayawardhana}{Nguyen et~al.}{2012}]{Nguyen2012}
Nguyen D.~C.,  Brandeker A.,  van Kerkwijk M.~H.,   Jayawardhana R.,  2012,
  ApJ, 745, 119

\bibitem[\protect\citeauthoryear{Nixon, King  \& Pringle}{Nixon
  et~al.}{2018}]{Nixon2018}
Nixon C.~J.,  King A.~R.,   Pringle J.~E.,  2018, MNRAS, 477, 3273

\bibitem[\protect\citeauthoryear{Pascucci, Cabrit, Edwards, Gorti, Gressel  \&
  Suzuki}{Pascucci et~al.}{2022}]{Pascucci2022}
Pascucci I.,  Cabrit S.,  Edwards S.,  Gorti U.,  Gressel O.,   Suzuki T.,
  2022, arXiv:2203.10068

\bibitem[\protect\citeauthoryear{Percy, Gryc, Wong  \& Herbst}{Percy
  et~al.}{2006}]{Percy2006}
Percy J.~R.,  Gryc W.~K.,  Wong J. C.~Y.,   Herbst W.,  2006, PASP, 118, 1390

\bibitem[\protect\citeauthoryear{Perraut, Gravity  \& Collaboration}{Perraut
  et~al.}{2021}]{Perraut2021}
Perraut K.,  Gravity  Collaboration 2021, \mn@doi [The 20.5th Cambridge
  Workshop on Cool Stars, Stellar Systems, and the Sun]
  {10.5281/zenodo.4749037}

\bibitem[\protect\citeauthoryear{Petrov, Gullbring, Ilyin, Gahm, Tuominen,
  Hackman  \& Loden}{Petrov et~al.}{1996}]{Petrov1996}
Petrov P.~P.,  Gullbring E.,  Ilyin I.,  Gahm G.~F.,  Tuominen I.,  Hackman T.,
    Loden K.,  1996, A\&A, 314, 821

\bibitem[\protect\citeauthoryear{Petrov, Zajtseva, Efimov, Duemmler, Ilyin,
  Tuominen  \& Shcherbakov}{Petrov et~al.}{1999}]{Petrov1999}
Petrov P.,  Zajtseva G.,  Efimov Y.,  Duemmler R.,  Ilyin I.~V.,  Tuominen I.,
   Shcherbakov V.~A.,  1999, A\&A, 341, 553

\bibitem[\protect\citeauthoryear{Petrov, Gahm, Stempels, Walter  \&
  Artemenko}{Petrov et~al.}{2011}]{Petrov2011}
Petrov P.,  Gahm G.~F.,  Stempels H.~C.,  Walter F.~M.,   Artemenko S.~A.,
  2011, A\&A, 535, 6

\bibitem[\protect\citeauthoryear{Petrov et~al.,}{Petrov
  et~al.}{2019}]{Petrov2019}
Petrov P.~P.,  et~al., 2019, MNRAS, 483, 132

\bibitem[\protect\citeauthoryear{Petrov, Romanova, Grankin, Artemenko, Babina
  \& Gorda}{Petrov et~al.}{2021}]{Petrov2021}
Petrov P.~P.,  Romanova M.~M.,  Grankin K.~N.,  Artemenko S.~A.,  Babina E.~V.,
    Gorda S.~Y.,  2021, MNRAS, 504, 871

\bibitem[\protect\citeauthoryear{Powell, Irwin, Bouvier  \& Clarke}{Powell
  et~al.}{2012}]{Powell2012}
Powell S.~L.,  Irwin M.,  Bouvier J.,   Clarke C.~J.,  2012, MNRAS, 426, 3315

\bibitem[\protect\citeauthoryear{Romanova \& Owocki}{Romanova \&
  Owocki}{2015}]{Romanova2015}
Romanova M.~M.,  Owocki S.~P.,  2015, SSRv, 191, 339

\bibitem[\protect\citeauthoryear{Romanova, Ustyugova, Koldoba  \&
  Lovelace}{Romanova et~al.}{2009}]{Romanova2009}
Romanova M.~M.,  Ustyugova G.~V.,  Koldoba A.~V.,   Lovelace R. V.~E.,  2009,
  MNRAS, 399, 1802

\bibitem[\protect\citeauthoryear{Romanova, Blinova, Ustyugova, Koldoba  \&
  Lovelace}{Romanova et~al.}{2018}]{Romanova2018}
Romanova M.~M.,  Blinova A.~A.,  Ustyugova G.~V.,  Koldoba A.~V.,   Lovelace R.
  V.~E.,  2018, NewA, 62, 94

\bibitem[\protect\citeauthoryear{S.~Lawrence}{S.~Lawrence}{1987}]{Lawrence1987}
S.~Lawrence M.~J.,  1987, Digital spectral analysis with applications.
Prentice-Hall

\bibitem[\protect\citeauthoryear{Shu, Najita, Ostriker, Wilkin, Ruden  \&
  Lizano}{Shu et~al.}{2016}]{Shu1994}
Shu F.,  Najita J.,  Ostriker E.,  Wilkin F.,  Ruden S.,   Lizano S.,  2016,
  ApJ, 820, 139

\bibitem[\protect\citeauthoryear{Skinner, Audard  \& Guedel}{Skinner
  et~al.}{2011}]{Skinner2011}
Skinner S.,  Audard M.,   Guedel M.,  2011, ApJ, 737, 19

\bibitem[\protect\citeauthoryear{Sousa et~al.,}{Sousa et~al.}{2021}]{Sousa2021}
Sousa A.~P.,  et~al., 2021, A\&A, 649, 68

\bibitem[\protect\citeauthoryear{St-Onge \& Bastien}{St-Onge \&
  Bastien}{2008}]{St-Onge2008}
St-Onge G.,  Bastien P.,  2008, ApJ, 674, 1032

\bibitem[\protect\citeauthoryear{Strassmeier, Bartus, Cutispoto  \&
  Rodono}{Strassmeier et~al.}{1997}]{Strassmeier1997}
Strassmeier K.~G.,  Bartus J.,  Cutispoto G.,   Rodono M.,  1997, A\&AS, 125,
  11

\bibitem[\protect\citeauthoryear{Tambovtseva, Grinin  \& Weigelt}{Tambovtseva
  et~al.}{2014}]{Tambovtseva2014}
Tambovtseva L.~V.,  Grinin V.~P.,   Weigelt G.,  2014, A\&A, 562, 104

\bibitem[\protect\citeauthoryear{Varniere \& Tagger}{Varniere \&
  Tagger}{2006}]{Varniere2006}
Varniere P.,  Tagger M.,  2006, A\&A, 446, 13

\bibitem[\protect\citeauthoryear{Xu, Herczeg, Johns-Krull  \& France}{Xu
  et~al.}{2021}]{Xu2021}
Xu Z.,  Herczeg G.,  Johns-Krull C.,   France C.,  2021, ApJ, 921, 181

\bibitem[\protect\citeauthoryear{Yu \& Lai}{Yu \& Lai}{2013}]{Yu2013}
Yu C.,  Lai D.,  2013, MNRAS, 429, 2748

\bibitem[\protect\citeauthoryear{Zajtseva}{Zajtseva}{2010}]{Zajtseva2010}
Zajtseva G.~V.,  2010, Ap, 53, 212

\bibitem[\protect\citeauthoryear{Zajtseva, Petrov, Ilyin, Duemler  \&
  Tuominen}{Zajtseva et~al.}{1996}]{Zajtseva1996}
Zajtseva G.,  Petrov P.,  Ilyin I.,  Duemler R.,   Tuominen I.,  1996, IBVS,
  4408, 1

\bibitem[\protect\citeauthoryear{Zanni \& Ferreira}{Zanni \&
  Ferreira}{2013}]{Zanni2013}
Zanni C.,  Ferreira J.,  2013, A\&A, 550, 99

\bibitem[\protect\citeauthoryear{Zhu, Hartmann, Calvet, Hernandez, Muzerolle
  \& Tannirkulam}{Zhu et~al.}{2007}]{Zhu2007}
Zhu Z.,  Hartmann L.,  Calvet N.,  Hernandez J.,  Muzerolle J.,   Tannirkulam
  A.-K.,  2007, ApJ, 669, 483

\makeatother
\end{thebibliography}

%\begin{thebibliography}{99}
%\end{thebibliography}

%%%%%%%%%%%%%%%%%%%%%%%%%%%%%%%%%%%%%%%%%%%%%%%%%%

%%%%%%%%%%%%%%%%% APPENDICES %%%%%%%%%%%%%%%%%%%%%

%\appendix

\newpage
\appendix
\section{Observation data}

\begin{table*}
\centering
\caption{RY Tau V-magnitudes for the dates of spectral observations in 2018-2022. The Heliocentric Julian Date (HJD) in the first column is followed by the site of the observation (CrAO -  Crimean Astrophysical Observatory, UrFU - Kourovka astronomical observatory of Ural Federal University), magnitude measured in the V-band, corresponding error and source (CrAO, AAVSO - American Association of Variable Star Observers, Int - value interpolated from AAVSO and CrAO phometry).}
\label{tab:photRY}
\begin{tabular}{lcccc} % four columns, alignment for each
\hline	
\hline
HJD-2400000& Site & V      & V error & Source of V   \\
\hline
58385.4814 & CrAO & 9.95  & 0.01 & CrAO \\
58387.4837 & CrAO & 9.93  & 0.10 & Int \\
58388.3744 & CrAO & 9.92  & 0.01 & CrAO \\
58389.3881 & CrAO & 9.80  & 0.01 & CrAO \\
58390.4398 & CrAO & 9.87  & 0.01 & CrAO \\
58410.4249 & CrAO & 9.46  & 0.01 & CrAO \\
58413.3276 & CrAO & 9.58  & 0.01 & CrAO \\
58414.4158 & CrAO & 9.59  & 0.01 & CrAO \\
58415.4108 & CrAO & 9.64  & 0.01 & CrAO \\
58508.1821 & CrAO & 9.84  & 0.01 & CrAO \\
58509.1849 & CrAO & 9.71  & 0.01 & CrAO \\
58511.3288 & CrAO & 9.80  & 0.10 & Int \\
58539.3164 & CrAO & 10.78 & 0.01 & CrAO \\
58567.2115 & CrAO & 10.78 & 0.10 & Int \\
58733.4181 & CrAO & 10.01 & 0.01 & CrAO \\
58734.4325 & CrAO & 9.86  & 0.01 & CrAO \\
58735.4190 & CrAO & 10.03 & 0.01 & CrAO \\
58736.4114 & CrAO & 10.02 & 0.01 & CrAO \\
58737.4201 & CrAO & 9.98  & 0.01 & CrAO \\
58765.4148 & CrAO & 9.74  & 0.01 & CrAO \\
58766.3949 & CrAO & 9.65  & 0.01 & CrAO \\
58767.4075 & CrAO & 9.60  & 0.01 & CrAO \\
58768.4256 & CrAO & 9.56  & 0.01 & CrAO \\
58793.4984 & CrAO & 9.80  & 0.01 & CrAO \\
58794.4122 & CrAO & 9.81  & 0.10 & Int \\
58795.3102 & CrAO & 9.82  & 0.01 & CrAO \\
58796.4116 & CrAO & 9.78  & 0.01 & CrAO \\
58822.1974 & CrAO & 10.07 & 0.01 & CrAO \\
58823.1794 & CrAO & 9.95  & 0.01 & CrAO \\
58824.1764 & CrAO & 9.83  & 0.01 & CrAO \\
58825.1786 & CrAO & 9.87  & 0.01 & CrAO \\
58826.3344 & CrAO & 9.96  & 0.01 & CrAO \\
58832.1747 & CrAO & 9.72  & 0.01 & CrAO \\
58834.1948 & CrAO & 9.74  & 0.01 & CrAO \\
58835.1767 & CrAO & 9.72  & 0.01 & CrAO \\
58836.1848 & CrAO & 9.72  & 0.01 & CrAO \\
58837.1782 & CrAO & 9.67  & 0.01 & CrAO \\
59094.4484 & CrAO & 10.21 & 0.01 & CrAO \\
59095.4479 & CrAO & 10.15 & 0.10 & Int \\
59096.4027 & CrAO & 10.10 & 0.01 & AAVSO \\
59118.4230 & CrAO & 10.09 & 0.10 & Int \\
59119.4668 & CrAO & 10.09 & 0.01 & CrAO \\
59120.3924 & CrAO & 10.10 & 0.01 & CrAO \\
59121.4180 & CrAO & 10.10 & 0.01 & CrAO \\
59123.4526 & CrAO & 10.24 & 0.01 & AAVSO \\
59150.4246 & CrAO & 9.47  & 0.01 & CrAO \\
59151.4660 & CrAO & 9.43  & 0.01 & CrAO \\ 
59151.5010 & UrFU & 9.43  & 0.01 & CrAO \\ 
59157.4042 & UrFU & 9.50  & 0.10 & Int \\
59174.3277 & CrAO & 9.70  & 0.10 & Int \\
59176.4235 & CrAO & 9.73  & 0.01 & CrAO \\
59178.4348 & CrAO & 9.77  & 0.01 & CrAO \\
59189.1893 & CrAO & 10.27 & 0.01 & CrAO \\
59190.2879 & CrAO & 10.03 & 0.01 & CrAO \\
59194.4090 & UrFU & 9.75  & 0.10 & Int \\
59195.4826 & UrFU & 9.73 & 0.10 & Int \\
59203.5020 & UrFU & 9.54 & 0.10 & Int \\
59217.1873 & CrAO & 9.66  & 0.01 & CrAO \\
&&&&\\
&&&&\\
&&&&\\
&&&&\\

%\hline
\end{tabular}
\quad
\begin{tabular}{ccccc}
\hline
\hline
HJD-2400000 & Site & V      & V error & Source of V   \\
\hline

59221.1563 & CrAO & 9.72  & 0.01 & CrAO \\
59251.1820 & CrAO & 10.01 & 0.01 & CrAO \\
59278.2196 & CrAO & 9.89  & 0.01 & CrAO \\
59279.2029 & CrAO & 9.94  & 0.01 & CrAO \\
59280.2914 & CrAO & 9.95  & 0.10 & Int \\
59288.2045 & CrAO & 9.94  & 0.01 & CrAO \\
59475.4156 & CrAO & 10.00 & 0.10 & Int \\
59477.4852 & CrAO & 9.96  & 0.01 & CrAO \\
59505.4315 & CrAO & 10.28 & 0.01 & CrAO \\
59508.3752 & CrAO & 10.28 & 0.01 & CrAO \\
59531.2225 & CrAO & 10.44 & 0.01 & CrAO \\
59532.2283 & CrAO & 10.65 & 0.01 & AAVSO \\
59533.2331 & CrAO & 10.67 & 0.10 & Int \\
59534.2230 & CrAO & 10.69 & 0.01 & CrAO \\
59544.2192 & CrAO & 10.37 & 0.01 & CrAO \\
59545.2085 & CrAO & 10.30 & 0.10 & Int \\
59586.3711 & CrAO & 10.10 & 0.10 & Int \\
59599.3919 & CrAO & 10.30 & 0.10 & Int \\
59624.1807 & CrAO & 10.38 & 0.01 & CrAO \\
59632.2225 & CrAO & 9.94  & 0.01 & CrAO \\
59652.2014 & CrAO & 10.03 & 0.01 & CrAO \\
59654.2259 & CrAO & 10.23 & 0.01 & CrAO \\
59663.2065 & CrAO & 9.76  & 0.01 & CrAO \\
59665.2201 & CrAO & 9.72  & 0.01 & CrAO \\
59666.2159 & CrAO & 9.77  & 0.01 & CrAO \\
&&&&\\
&&&&\\
&&&&\\
&&&&\\
&&&&\\
&&&&\\
&&&&\\
&&&&\\
&&&&\\
&&&&\\
&&&&\\
&&&&\\
&&&&\\
&&&&\\
&&&&\\
&&&&\\
&&&&\\
&&&&\\
&&&&\\
&&&&\\
&&&&\\
&&&&\\
&&&&\\
&&&&\\
&&&&\\
&&&&\\
&&&&\\
&&&&\\
&&&&\\
&&&&\\
&&&&\\
&&&&\\
&&&&\\
&&&&\\
&&&&\\
&&&&\\
&&&&\\
%\hline
\end{tabular}
\end{table*}

\begin{table*}
\centering
\caption{SU Aur V-magnitudes for the dates of spectral observations in 2018-2022. The Heliocentric Julian Date (HJD) in the first column is followed by the site of the observation (CrAO -  Crimean Astrophysical Observatory, UrFU - Kourovka astronomical observatory of Ural Federal University), magnitude measured in the V-band, corresponding error and source (CrAO, AAVSO - American Association of Variable Star Observers, Int - value interpolated from AAVSO and CrAO phometry).}
\label{tab:photSU}
\begin{tabular}{lcccc} % four columns, alignment for each
\hline	
\hline
HJD-2400000& Site & V      & V error & Source of V   \\
\hline
58385.5254 & CrAO &  9.35 & 0.01 & CrAO \\
58387.5306 & CrAO &  9.55 & 0.01 & CrAO \\
58388.4419 & CrAO &  9.33 & 0.01 & CrAO \\
58389.4523 & CrAO &  9.36 & 0.01 & CrAO \\
58390.5045 & CrAO &  9.40 & 0.01 & CrAO \\
58410.4894 & CrAO &  9.34 & 0.01 & CrAO \\
58413.3925 & CrAO &  9.32 & 0.01 & CrAO \\
58414.4808 & CrAO &  9.35 & 0.01 & CrAO \\
58415.4753 & CrAO &  9.34 & 0.01 & CrAO \\
58508.2357 & CrAO &  9.53 & 0.01 & CrAO \\
58509.2595 & CrAO &  9.50 & 0.01 & CrAO \\
58511.3531 & CrAO &  9.40 & 0.10 & Int \\
58521.3400 & UrFU &  9.35 & 0.10 & Int \\
58535.2368 & UrFU &  9.51 & 0.01 & CrAO \\
58539.2519 & CrAO &  9.44 & 0.01 & CrAO \\
58559.2333 & UrFU &  9.62 & 0.02 & AAVSO \\
58567.2619 & CrAO &  9.60 & 0.10 & Int \\
58568.2508 & CrAO &  9.52 & 0.01 & CrAO \\
58733.4841 & CrAO &  9.40 & 0.01 & CrAO \\
58734.4970 & CrAO &  9.44 & 0.01 & CrAO \\
58735.4834 & CrAO &  9.34 & 0.01 & CrAO \\
58736.4770 & CrAO &  9.26 & 0.01 & CrAO \\
58737.4841 & CrAO &  9.27 & 0.01 & CrAO \\
58765.4834 & CrAO &  9.32 & 0.01 & CrAO \\
58766.4600 & CrAO &  9.33 & 0.01 & CrAO \\
58767.4518 & CrAO &  9.29 & 0.01 & CrAO \\
58768.4906 & CrAO &  9.34 & 0.01 & CrAO \\
58793.4281 & CrAO &  9.27 & 0.10 & Int \\
58794.4564 & CrAO &  9.28 & 0.01 & CrAO \\
58795.3754 & CrAO &  9.29 & 0.01 & CrAO \\
58796.4773 & CrAO &  9.24 & 0.01 & CrAO \\
58822.2870 & CrAO &  9.34 & 0.01 & CrAO \\
58823.2465 & CrAO &  9.25 & 0.01 & CrAO \\
58824.2414 & CrAO &  9.29 & 0.01 & CrAO \\
58825.2342 & CrAO &  9.28 & 0.01 & CrAO \\
58826.3785 & CrAO &  9.27 & 0.01 & CrAO \\
58832.2412 & CrAO &  9.27 & 0.01 & CrAO \\
58834.2621 & CrAO &  9.26 & 0.01 & CrAO \\
58835.2633 & CrAO &  9.24 & 0.01 & CrAO \\
58836.2718 & CrAO &  9.23 & 0.01 & CrAO \\
58837.2858 & CrAO &  9.21 & 0.01 & CrAO \\
58847.3812 & UrFU &  9.32 & 0.10 & Int \\
58851.2413 & CrAO &  9.36 & 0.01 & CrAO \\
58862.3010 & CrAO &  9.26 & 0.01 & CrAO \\
58864.2685 & CrAO &  9.24 & 0.01 & CrAO \\
58865.2706 & CrAO &  9.25 & 0.01 & CrAO \\
58877.2560 & UrFU  &  9.27 & 0.01 & Int \\
58889.3070 & CrAO &  9.41 & 0.01 & CrAO \\
58890.2680 & CrAO &  9.45 & 0.01 & CrAO \\
58914.2987 & UrFU  &  9.55 & 0.10 & Int \\
58916.2893 & CrAO &  9.51 & 0.01 & CrAO \\
58918.2852 & CrAO &  9.37 & 0.01 & CrAO \\
58923.2899 & CrAO &  9.35 & 0.01 & CrAO \\
58925.2692 & CrAO &  9.55 & 0.01 & CrAO \\
58926.2454 & CrAO &  9.63 & 0.01 & CrAO \\
58932.2576 & UrFU &  9.60 & 0.10 & Int \\
58936.2847 & UrFU &  9.60 & 0.10 & Int \\
59094.5135 & CrAO &  9.45 & 0.01 & CrAO \\
&&&&\\
&&&&\\
&&&&\\
&&&&\\
\end{tabular}
\quad
\begin{tabular}{ccccc}
\hline
\hline
HJD-2400000 & Site & V    & V error & Source of V   \\
\hline
59095.5174 & CrAO &  9.47 & 0.10 & Int \\
59096.4822 & CrAO &  9.50 & 0.10 & Int \\
59119.5152 & CrAO &  9.86 & 0.01 & CrAO \\
59120.4622 & CrAO & 10.00 & 0.01 & CrAO \\
59121.4901 & CrAO & 10.03 & 0.01 & CrAO \\
59150.5046 & CrAO &  9.50 & 0.01 & CrAO \\
59151.4993 & UrFU &  9.56 & 0.01 & CrAO \\
59151.5310 & CrAO &  9.56 & 0.01 & CrAO \\
59174.3962 & CrAO &  9.64 & 0.10 & Int \\
59176.4926 & CrAO &  9.65 & 0.01 & CrAO \\
59189.2371 & CrAO &  9.58 & 0.01 & CrAO \\
59190.3823 & CrAO &  9.64 & 0.01 & CrAO \\
59217.2541 & CrAO &  9.68 & 0.01 & CrAO \\
59218.2252 & CrAO &  9.68 & 0.01 & CrAO \\
59219.2124 & CrAO &  9.70 & 0.01 & CrAO \\
59221.2569 & CrAO &  9.83 & 0.01 & CrAO \\
59249.2494 & CrAO &  9.40 & 0.01 & CrAO \\
59257.3505 & UrFU & 9.33 & 0.10 & AAVSO \\
59275.3060 & UrFU & 9.36 & 0.10 & Int \\
59279.2716 & CrAO &  9.34 & 0.01 & CrAO \\
59287.2796 & UrFU & 9.34 & 0.10 & Int \\
59288.2700 & CrAO & 9.35 & 0.01 & CrAO \\
59288.2886 & UrFU & 9.35 & 0.01 & CrAO \\
59296.2526 & UrFU & 10.03 & 0.10 & AAVSO \\
59318.2247 & UrFU &  9.90 & 0.10 & Int \\
59477.5286 & CrAO &  9.72 & 0.01 & CrAO \\
59505.4855 & CrAO &  9.55 & 0.01 & CrAO \\
59533.2996 & CrAO &  9.50 & 0.10 & Int \\
59544.3023 & CrAO &  9.73 & 0.01 & CrAO \\
59586.4188 & CrAO &  9.60 & 0.10 & Int \\
59598.2719 & CrAO &  9.58 & 0.01 & CrAO \\
59599.4372 & CrAO &  9.58 & 0.10 & Int \\
59624.2683 & CrAO &  9.53 & 0.01 & CrAO \\
59628.2886 & CrAO &  9.49 & 0.01 & CrAO \\
59630.2814 & CrAO &  9.45 & 0.10 & Int \\
59632.2727 & CrAO &  9.38 & 0.01 & CrAO \\
59652.2676 & CrAO &  9.44 & 0.01 & CrAO \\
59663.2538 & CrAO &  9.43 & 0.01 & CrAO \\
59665.2656 & CrAO &  9.43 & 0.01 & CrAO \\
59666.2610 & CrAO &  9.55 & 0.01 & CrAO \\
59667.2560 & CrAO &  9.53 & 0.01 & CrAO \\
&&&&\\
&&&&\\
&&&&\\
&&&&\\
&&&&\\
&&&&\\
&&&&\\
&&&&\\
&&&&\\
&&&&\\
&&&&\\
&&&&\\
&&&&\\
&&&&\\
&&&&\\
&&&&\\
&&&&\\
&&&&\\
&&&&\\
&&&&\\
&&&&\\
\end{tabular}
\end{table*}


%%%%%%%%%%%%%%%%%%%%%%%%%%%%%%%%%%%%%%%%%%%%%%%%%%

% Don't change these lines
\bsp	% typesetting comment
\label{lastpage}
\end{document}